\documentstyle[aps,prd,eqsecnum,preprint]{revtex}
\begin{document}
\input{epsf}
\draft 
\title {Estimation of Parameters of Gravitational Wave Signals from Coalescing 
Binaries.}
\author{R. Balasubramanian and S. V. Dhurandhar}
\address{Inter-University Centre for Astronomy and Astrophysics, \\
Post Bag 4, Ganeshkhind, Pune 411 007, India.}
\date{\today}
\maketitle
\begin{abstract}
With the on going construction of several large and medium scale laser
interferometric gravitational wave antennas around the globe, the
problem of the  detection of gravitational waves has acquired great
impetus. Since gravitational  wave signals from astrophysical
sources  are expected to be weak, despite the state of 
the art technology being 
employed, the development of optimal signal extraction techniques 
and the consequent accurate determination of the parameters of the signal is of major 
importance.  Coalescing binary systems are one of the most promising sources of
gravitational waves. One reason is that such sources are easier
to model and thus one can design detection strategies 
tuned to such signals. A lot of attention has been devoted in the 
literature studying such techniques and most of the work has 
revolved around matched filtering and  maximum likelihood 
estimation.  

In a previous work, Monte Carlo simulations were carried out  
of the detection process using matched filtering for the initial LIGO/VIRGO 
configuration for the first post-Newtonian corrected coalescing binary waveform. 
We had compared the results of our simulations with available 
estimates, obtained from covariance matrix considerations, of the errors in the 
determination of the parameters. Our results
showed that the covariance matrix underestimates, by over a factor of two,
the actual errors in the estimation of parameters even when the signal-to-noise  
ratio (SNR) 
is as high as 10. Sources having SNR higher than 10 are expected to be few and 
hence this issue is of major concern. 

In this paper we probe the question as to why the Monte Carlo simulations give 
such high errors as opposed to those obtained via the covariance matrix.
We present, a {\em computationally
viable} statistical model of the distribution, of the maximum likelihood 
estimates (MLE), of  the parameters. This model reproduces the essential
features of the Monte Carlo simulations, thereby explaining 
the large root mean square errors in the estimates, obtained in numerical
experiments. The chief reason for the large errors seems to be the fact
that the probability 
distribution of the estimated parameters is multimodal. Though only
the central peak (corresponding to the actual values of the parameters)
is dominant, the subsidary  peaks occur `far' away thereby contributing to
large variances. We  suggest  that the variance or the standard deviation of an
estimated parameter may not provide the best measure of the error, for the kind 
of situation we encounter 
here. We therefore propose another criterion by which the MLE  
should be judged.

In order to illustrate the model we have considered the Newtonian as
well as the first post-Newtonian corrected waveform.
We have assumed  Gaussian noise, 
with a power spectrum typical of the LIGO/VIRGO type of detectors.
The model we have used, however, is quite general,
and robust, and will be relevant to many other parameter estimation problems.
\end{abstract}

\section{Introduction}
Large scale laser interferometric detectors of gravitational waves, namely, the LIGO 
\cite{LIGO} and VIRGO \cite{VIRGO} and the medium scale detectors, GEO
and TAMA are expected to be operational by the turn of this century. 
Compact coalescing binary systems of blackholes and/or neutron stars 
are relatively `clean' systems to model during their inspiral and their inspiral  
waveform can be predicted with a fair degree 
of reliability. This makes them the most promising sources for broad band detectors, 
and in particular, the upcoming interferometric detectors cited above. 
Binary systems are also valuable sources of astrophysical information as 
one can probe the universe up to cosmological distances. For instance,
statistical analysis of several binary coalescences enables the estimation
of the Hubble constant to an accuracy better than 10\% \cite{SCH86,Mar93,Fin96}. 
Events that produce high signal-to-noise ratio can be potentially used
to observe such non-linear effects, as gravitational wave
tails, and to put general relativity into test in the strongly non-linear
regime \cite{BS95}.
Due to the weak coupling of gravitational radiation with matter the 
signal waveform has a very low amplitude and will not stand above the detector noise.
In addition to the on-going efforts to reduce the noise, and hence increase the
sensitivity of the detector, a considerable amount of research activity
has gone into the development of efficient and robust data analysis techniques 
to extract signals buried in very noisy data.
For a review on gravitational waves from compact objects and their 
detection see Thorne \cite{Th95a,Th95b}.

Various data analysis schemes have been proposed for the detection of the 
`chirp' waveform from such systems. Among  them the 
technique of matched filtering is the most promising \cite{Th87,HEL,SCH89}. 
Briefly, this technique involves correlating the detector output 
with a set of templates, each of which is tuned to
detect the signal with a particular set of parameters. In order to obtain a high 
value for the correlation the signal waveform should be known 
to a high level of accuracy. The matched filtering technique is very sensitive to the 
{\it phase} of the signal and even if the template and the signal mismatch by even 
half a cycle the correlation integral is drastically reduced. 

The fully general relativistic
waveform from a coalescing binary system of stars is as yet unavailable.
In the absence of such an exact solution, there have been
efforts to find solutions perturbatively. Most of the work  in
this area strives towards computing the waveform correct to a high degree of
accuracy so that the theoretical templates based on this will
obtain the maximum possible SNR.

The signal is said to be detected, if the maximum value of the correlation over 
all the parameters of the signal crosses a preassigned threshold which has been set 
by the false alarm one is prepared to tolerate. Once the signal is
detected, the maximum likelihood estimates (MLE)
 of the  parameters of the binary are  those of 
the template with which the maximum correlation is obtained. The errors 
involved in such an estimation have been worked out by several
authors \cite{BS95,Fin92,FC93,BS94,Kr,CF94,KLM93,PW95}, 
for the case of `high' SNR and for the Newtonian and post-Newtonian 
waveforms using a single and a network of detectors. 
In \cite{BSD96} exhaustive Monte Carlo numerical simulations were carried out to 
compute the errors in the
estimation of parameters and covariances among them,
for the case of the initial LIGO configuration taking 
into account only the first post-Newtonian corrections and 
assuming circular orbits. It was found that the errors as obtained from the 
simulations were larger by a factor of two or more from the errors as computed via 
the covariance matrix at astrophysically relevant SNRs. 
The discrepancy disappears as the SNR 
increases beyond certain value - typically 15 in the Newtonian case and 25 in the 
post-Newtonian case. 
The comparision with other less stringent 
lower bounds has also not resolved the discrepancy.
Nicholson and Vecchio \cite{NV97} have recently computed Bayesian 
bounds such as the  Weiss-Wainstein and Ziv-Zakai, for the case of
Newtonian signals. They conclude that though these bounds are tighter
than the Cramer-Rao bound, the numerical errors are still much larger
than the computed lower bounds. 

In this paper we explain the discrepancy between the Monte Carlo simulations and the 
results obtained from the covariance matrix. We demonstrate that the
probability distribution of the estimated parameters cannot be got
from `local' methods such as the ones used earlier \cite{Fin92}.
Since our main purpose in this paper is to demonstrate the
validity of the statistical model,
we initially use the simplified Newtonian model of the waveform.
In the Newtonian case there are fewer parameters of the signal that 
one has to deal with,  making the investigations simpler, analytically as well as
computationally. We then specialize to the first post-Newtonian
case. Here, although the discrepancy is larger, the problem is similar
at a qualitative level.

Following Finn \cite{Fin92} we obtain an 
equation which relates the estimated parameters to the actual parameters of the 
signal for a given noise realisation. This equation is non-linear. Finn linearises 
the equation and obtains the errors in terms of the covariance matrix. Here we do not 
linearise the equation. We find that the equation has multiple solutions for the parameters 
in a certain sense and it is this multiplicity of roots forming islands in the parameter 
space which contributes significantly to the errors. Thus the problem is of 
a global nature and any local approximation will be inadequate in explaining away the 
errors. Moreover we suggest that the variance/covariance of the parameters is not a 
proper measure of the errors in this case. We therefore propose  a new criterion
by which the MLE should be judged.  

The paper is organized as follows.   
In section \ref{waveform} we briefly describe the gravitational wave signal
waveform and the  parameters on which it depends, namely,
the amplitude, the time of arrival, the 
phase at arrival and the `chirp times', which characterizes the time for
the binary to evolve from some fiducial time to the actual merger.
These parameters are 
found to be very convenient when we carry out Monte Carlo simulations. 
It turns out that the covariance matrix  is independent of these parameters 
and hence it is sufficient to carry out the simulations only  for a particular 
set of parameters. Further in this section we also describe the characterstics of 
the noise that we assume will be present in the detector and briefly
review the matched filtering technique. In section \ref{MCres} we
present the results of the Monte Carlo simulations for the Newtonian case.
We carry out transformations of the parameter space which bring out the
chief features of the distribution of the estimated parameters.
We show that the estimated parameters do not lie in a simply connected
region around the signal parameters, but
instead are distributed in multiple  `islands' in the parameter space
considered.
We next present a geometric representation of our statistical
model. We then apply this model to the Newtonian chirp waveform, 
and compare the model with the Monte Carlo simulations.
In section \ref{PNwave} we deal with the post-Newtonian waveform.
Finally in section~\ref{sec_con}  we 
summarise our results. We propose an alternative measure for the error
which performs reasonably better 
than the variance as a measure of the error.

\section{The Signal and the Noise}
\label{waveform}

\subsection {The Chirp Signal}
\label{chirp}
When constructing templates for on-line detection,
it is sufficient to work with the so called {\it restricted\,} post-Newtonian
gravitational waveform. In this approximation 
the post-Newtonian corrections are incorporated only in the phase of the 
waveform, while ignoring corresponding corrections to the amplitude
\cite {3mn}. Consequently, the restricted post-Newtonian waveforms
only contain the dominant frequency equal to twice the orbital frequency
of the binary computed up to the relevant order.
In the restricted post-Newtonian approximation the gravitational
waves from a binary system of stars, modeled as point masses orbiting
about each other in a circular orbit, induce a strain $s(t)$ at the
detector given by
\begin {equation}
s(t) = {\cal A} (\pi f(t) )^{2/3} \cos \left [\varphi (t) + \Phi\right],
\label {wave}
\end {equation}
where $f(t)$ is the instantaneous gravitational wave frequency,
the constant $\cal A$ involves the distance to the binary, its reduced
and total mass, and the antenna pattern of the detector \cite{Th87}
and $\Phi$ is the initial phase of the wave at some fiducial time
$t=t_s$. 
The phase of the waveform $\varphi (t)$ contains several pieces 
corresponding to different post-Newtonian contributions which
can be schematically written as
\begin {equation}
\varphi(t) = \varphi_0(t) + \varphi_1(t) + \varphi_{1.5}(t) + \ldots.
\label {phase}
\end {equation}
The evolution of the phase depends on the masses of the two components
of the binary, characterised by the reduced mass, $\mu$ and the total
mass, $M$ of the system.  
Here $\varphi_0(t)$ is the dominant Newtonian part of the phase 
and $\varphi_n(t)$ represents the $n$th order post-Newtonian
correction to it. The Newtonian part of the phase is sensitive only to
a particular combination of the masses of the two stars, frequently
characterised by its `chirp mass', ${\cal M} = \mu^{3/5}M^{2/5}$.
We give below the waveform correct to the first
post-Newtonian order:
\begin {eqnarray}
\varphi_0 (t) &=& {16 \pi f_s\tau_0 \over 5} 
\left [ 1 - \left ({f\over f_s}\right )^{-5/3} \right] \nonumber,\\
\varphi_1(t)&=& 4 \pi f_s\tau_1 \left [ 1 - \left ( {f\over f_s} \right )^{-1} \right ]
\label {phaseN}
\end {eqnarray}
where $f(t)$ is   given implicitly by,
\begin {equation}
t - t_s= \tau_0
\left [ 1 - \left ( {f \over f_s} \right )^{-8/3} \right ]+
\tau_1\left[1 - \left ( {f \over f_s} \right )^{-2} \right]
\label {frequencyN},
\end {equation}
where $\tau_0$ and $\tau_1$ are constants having dimensions
of time given by
\begin {eqnarray}
\tau_0 &=& {5 \over 256} {\cal M}^{-5/3} (\pi f_s)^{-8/3},\nonumber\\
\tau_1 &=&  {5 \over 192\mu (\pi f_s)^2} \left ({743\over 336} + {11\over 4} \eta
\right ),
\label{NCT}
\end {eqnarray}
where $\eta=\mu/M$, and $f_s$  is the instantaneous gravitational 
wave frequency of the signal at $t=t_s.$ 
The time elapsed starting from an epoch when the 
gravitational wave frequency is $f_s$ till the epoch when
it becomes infinite will be referred to as 
the {\it chirp time} of the signal.
In the quadrupole approximation $\tau_0$ is the chirp time
whereas it is $\tau_0+\tau_1$ for the first post-Newtonian case \cite{Sat94}.
The Newtonian part of the phase is characterised by three parameters:
(i) the {\it time of arrival} $t_s$ 
when the signal first becomes {\it visible} in the detector, 
(ii) the {\it phase} $\Phi$ of the signal at the time of arrival 
and (iii) the  chirp mass.  
At this level of approximation two coalescing binary signals of 
the same chirp mass but 
of different sets of individual masses would be degenerate and thus exhibit
exactly the same time evolution.  This
degeneracy is removed when post-Newtonian corrections are included.
The parameters $t_s$ and $\Phi$ are {\it kinematical} that fix the origin of the 
measurement of time and phase, respectively, while the Newtonian and the 
post-Newtonian chirp times 
are {\it dynamical} parameters in the sense that they dictate the
evolution of the phase and the amplitude of the signal.

The parameters $\tau_0$, $\tau_1$ and $t_s$ have the dimensions of
time. We convert them to dimensionless parameters by multiplying them
with $2\pi f_s$. Thus we have the parameter set, 
$$ 
\bbox{\mu} \equiv 
\{\mu^0,\mu^1,\mu^2,\mu^3,\mu^4\} \equiv \{A,2\pi f_st_s, \Phi, 2\pi
f_s\tau_0, 2\pi f_s\tau_1\}.
$$

In the stationary phase approximation 
the Fourier transform of the restricted second-post-Newtonian
chirp waveform for positive frequencies is given by \cite
{Th87,SD91,FC93,CF94}.
\begin {equation}
\tilde s (f) = A {\cal N} f^{-7/6} \exp \left [i\sum_{j=1}^4\chi_j(f)\mu^j
- i {\pi \over 4} \right ] \ ,   
\label {FT}
\end {equation} 
where 
$A$, is the amplitude parameter depending on the distance to the
binary, as well as the chirp mass and $\cal N$ is a normalization
constant to be fixed by means of a scalar product to be introduced later, and 
\begin {eqnarray}
\label {eqs1}
\chi_1 & = & {f\over f_s}, \nonumber\\
\chi_2 & = & -1, \nonumber\\
\chi_3 & = & {f\over f_s}  -{ 8 \over 5}+ {3\over 5}
\left ( {f\over f_s} \right )^{-5/3},\nonumber\\
\chi_4 & = & {f\over f_s} - 2 + {f_s\over f}.
\end {eqnarray} 
For $f<0$ the Fourier transform is computed using the identity $\tilde
s(-f) = \tilde s^*(f)$ obeyed by real functions $s(t).$ 

\subsection{The noise}
\label{noise}
The output of a gravitational wave detector such as the 
LIGO, will comprise of data segments, each of 
duration $T$ seconds, uniformly sampled with a sampling interval of 
$\Delta$, giving the number of samples in a single data train to be 
$N = T/\Delta$. Each data train can be considered as a $N$-tuple 
${\bf x} \equiv \{x^0,x^1,\ldots,x^{N-1}\}$,  $x^a$ 
being the  value of the output of the 
detector at time $a\Delta$. The set of all such $N$-tuples constitutes an
$N$-dimensional  vector space $\cal V$  where the addition of two vectors is 
accomplished by the addition of corresponding time samples. 
For later convenience we allow each sample to take complex values. A natural
basis for this vector space is  the {\em time basis}\, ${\bf e}_r^a = 
\delta^a_r$ where $r$ and $a$ are the vector and component indices 
respectively. Another basis which we shall use extensively is the Fourier 
basis.

A gravitational wave  signal from a coalescing binary system 
can  be characterised by a set of parameters 
$\mu^a, a = 0,1, ...,m-1$ belonging to some 
open set of the $m$-dimensional real space $R^m$.
The set of such signals
${\bf s}(t; \bbox{\mu})$ constitutes a $m$-dimensional manifold $\cal S$ which is 
embedded in the vector space $\cal V$. Note that Greek characters in
boldface denote the full collection of parameters characterising the signal. 
The parameters of the binary can be regarded as
coordinates on the manifold. The basic problem of signal analysis is thus to
 determine whether the detector output vector $\bf x$ is  
the sum of a signal vector and a noise vector, ${\bf x} = {\bf s} + {\bf n}$, 
or just the noise vector, ${\bf x} = {\bf n}$, and furthermore to identify which
particular signal vector, among all possible. The latter is relevant to this paper 
where we are interested in estimating the parameters and also the errors made 
in such a estimation. Errors in the estimation arise because the noise contaminates 
the data. 
 
The noise in ground based laser interferometric detectors will have,
 in general, both a Gaussian and a non-Gaussian component. The main 
sources for the Gaussian component are the shot noise due to photon
 counting, the thermal noise in 
the mirror suspensions alongwith the mirror itself and seismic noise. The
non-Gaussian component can be contributed by numerous sources like sudden strain
releases in the mirror suspension wires or even if lightning strikes.  
 It should be 
possible to remove most of the non-Gaussian component by using environmental
 monitors
and looking for coincidence among detectors located at widely separated sites.
It is, therefore, assumed usually that the detector noise will be a Gaussian
random process.
Over a time scale of hours, it can also be assumed to be stationary.

The power spectral density of the Gaussian noise component
rises very steeply towards the low frequency end due to seismic
effects. At the high frequency
end it is dominated by photon shot noise which leads to a rise towards higher
frequencies.
Thus the data will have to be bandpassed with a low frequency seismic
cutoff, $f_s$, and a high frequency cutoff, $f_c$.  We use the power spectral density
expected for the initial LIGO as given in \cite{LIGO}. Accordingly, we choose 
$f_s = 40$ Hz and  $ f_c = 800$ Hz.

In the absence of the signal the output will contain 
only noise drawn from a stochastic process which can be described by a 
probability distribution on the vector space $\cal V$. We assume that the noise 
has its mean zero, or that is, $\overline{n^a} = 0$, where the overbar
denotes an ensemble average. 
Then the covariance matrix of the noise ${\cal C}^{ab}$ is defined as, 
\begin{equation}
{\cal C}^{ab} = \overline{n^a n^b}.
\end{equation}
If the noise is assumed to be stationary and ergodic then 
there exists a noise autocorrelation function $K(t)$ such that ${\cal C}^{ab} = 
K(|a-b|\Delta)$. In the Fourier basis it can be shown that the components 
of the noise vector are statistically  independent \cite{HEL} and
    the covariance matrix in the Fourier
 basis will contain only diagonal terms whose values will be strictly positive:
 $\tilde{\cal C}^{aa} = \overline{\tilde n^a\tilde n^{*a}}$.
 This implies that the covariance matrix 
has strictly positive eigenvalues. The diagonal elements of this
matrix $\tilde {\cal C}^{aa}$ constitute the discrete representation of the power 
spectrum of the noise $S_n(f)$. 

Gaussian noise can be described by the distribution,
\begin{equation}
\label{ndis}
p_0 ({\bf n}) = M_n{\exp\left[
	-\frac{1}{2}\sum\limits_{a,b=0}^{N-1}{[{\cal C}^{-1}]_{ab} n^a n^{b}}\right]},
\end{equation}
where $M_n$ is a normalization constant given by,
$$
M_n = { \left[\ (2\pi)^N \det\left[{\cal C}^{ab}\right]\ \right]^{-1/2}}.
$$
Equivalently in the Fourier domain this can be written as,
\begin{eqnarray}
p_0({\bf n}) &=& M_n {\exp\left[
	-\frac{1}{2}\sum\limits_{a,b=0}^{N-1}{[\tilde {\cal C}^{-1}]_{ab} \tilde n^a 
\tilde n^{b*}}\right]}\nonumber\\
&=& M_n{\exp\left[
	-\frac{1}{2}\sum\limits_{a=0}^{N-1}{{\tilde n^a 
\tilde n^{a*}}/{\tilde {\cal C}^{aa}}}\right]},
\end{eqnarray}
where in the last step we have used the diagonal property of the matrix $\tilde {\cal C}^{ab}$
which implies that $[\tilde {\cal C}^{-1}]_{aa} = 1/\tilde {\cal C}^{aa}$.

In the presence of the signal ${\bf s} (\bbox{\check\mu})$ the above probability density 
function (pdf) gets modified but in a very simple way since we have assumed that the noise 
is additive. We have,
\begin{equation}
p_1 ({\bf x}) = p_0 ({\bf x} - {\bf s}(\bbox{\check\mu})),
\end{equation} 
where $p_1 ({\bf x})$ is the pdf of ${\bf x}$  when the signal ${\bf s} (\bbox{\check\mu})$ 
is present in the data stream.

In this paper we shall assume the noise to have a power spectrum
consistent with the initial LIGO instrument. We use the fit to the
noise curve as given in \cite{CF94}:
\begin{equation}
\label{psd}
S_n(f) = S_0 \left[\left(\frac{f}{200}\right)^{-4} + 2\left(1+\frac{f}{200}\right)^2\right].
\end{equation}
The value of $S_0$ will not concern us since what matters is only the
ratio of the signal amplitude to that of the noise, in other words the
SNR. We shall set the value of $S_0$ to be unity and accordingly  
adjust the amplitude of
the signal to get the required SNR.

\subsection{The Matched Filter}
\label{mf}

In the absence of any prior information it must be assumed that all the parameter values, 
within their respective range, are equally likely to occur. In such a case, 
the method of maximum liklihood can be used. When the  noise is  a stationary Gaussian 
random process, the method of MLE reduces to the  
so called matched filtering technique. Matched filtering involves correlating the detector 
output with a bank of matched filters each of which corresponds to the signal waveform 
for a fixed set of parameters. To this end, we  define a scalar product on $\cal V$. 
In the continuum limit the scalar product between two vectors ${\bf x}$ and ${\bf y}$ is 
given by,
\begin{equation}
\label{scal}
\left\langle{\bf x},{\bf y}\right\rangle   = \int_{0}^{\infty}df\,{1\over S_n(f)}\, 
 \left( \, \tilde{x}(f)\, \tilde{y}^{\ast} (f)\, + \,\tilde{x}^{\ast}(f)
 \,\tilde{y}(f)\, \right) \; ,
 \end{equation}
 where, the Hermitian property of the Fourier transform of a real function
 has been used
to restrict the domain of integration to positive frequencies. $S_n(f)$ is
 the power spectral density of the noise. The Fourier domain is convenient since stationarity 
of the noise has been assumed. The norm of a vector ${\bf z}$ will be denoted by 
$\|{\bf z}\| = \left\langle{\bf z},{\bf z}\right\rangle^{1/2}$.
In eqn.~(\ref{FT}) we had left $\cal N$ undefined. We choose the
value of  $\cal N$ such that $\|{\bf s}(\bbox{\mu})\| = A$.
From the definition of the scalar product in eqn.~(\ref{scal}) and
from eqn.~(\ref{FT}) it
follows that,
\begin{equation}
\label{normdef}
\frac{1}{{\cal N}^2} = 2\int\limits_0^\infty \frac{f^{-7/3}}{S_n(f)}df.
\end{equation} 
The integrand in the above equation, $I(f)$ is plotted in
Fig.~\ref{fig1}. 
This function peaks at around $f=135$Hz and a major
contribution to the integral comes from a region around $135$Hz.

We shall use normalized  matched filters, that is,we set 
${\bf h}(\mu^j) =  {\bf s}(\bbox{\mu})/A$. Henceforth we shall use the
symbols $a,b,\ldots$ to indicate indices whose range of values
includes $0$ which as a index denotes the amplitude parameter, {\em
i.e. $\mu^0 = A$}. Indices
$i,j,\ldots$ will not include the amplitude parameter and will never
take the value $0$.  

The output called the correlation $c(\mu^j)$ of the 
matched filter with parameters $\mu^j$ is then just the scalar product given by,
\begin{equation}
c(\mu^j) = \left\langle{\bf x},{\bf h}(\mu^j)\right\rangle.
\label{cor}
\end{equation}
Given the data vector ${\bf x}$, the correlation is computed for the entire 
feasible range of parameters, continuously over the kinematic parameters $t_s$ and 
$\Phi$ and for discrete values of the dynamical parameters $\tau_0$
and $\tau_1$. The filters 
corresponding to the discrete values of $\tau_0$ and $\tau_1$
 constitute the filter bank. The 
maximum of $c(\mu^j)$ is computed and compared with a preassigned threshold 
determined from the false alarm probability. A detection is announced   
if the maximum of $c(\mu^j)$ crosses the threshold.  The parameters $\hat \mu^j$ 
for which the correlation is 
maximised are the MLE of the parameters of the signal. 
However, these will in general differ from the actual signal parameters 
$\check \mu^j$ due to the presence of noise.  The difference 
$\Delta \mu^j = \check \mu^j - \hat \mu^j$  
is the error made in estimating the parameters.
When the signal is weak as compared to 
the noise (low SNR), the estimated parameters in general will differ by a large margin 
from the true ones, while in the limit of high SNR, the two will 
almost coincide. Thus in general $\Delta \mu^j$ is not small.

 Let the data vector be 
\begin{equation}
{\bf x} = \check A {\bf h}(\check \mu^j) + {\bf n},
\end{equation}
where $\check A$ is the amplitude of the signal and $\check \mu^j$ are the
remaining parameters.  We have separated the amplitude as it occurs in the 
signal as just a scale factor multiplying the signal. 
The SNR is defined as,
\begin{equation}
\label{SNR}
\rho =  \left\langle{\bf s}(\check{\bbox{\mu}}),{\bf 
s}(\check{\bbox{\mu}})\right\rangle^{1/2} = {\check A}.
\end{equation}
From eqs.~(\ref{FT}) and (\ref{scal}) we see that the SNR is equal
to the amplitude parameter of the signal, $\check A$.

When we consider an ensemble of noise realisations, the $c(\mu^j)$ becomes a 
random variable. For a fixed set of parameters $\mu^j$, $c(\mu^j)$ is
a Gaussian random variable since it is a linear function of the noise
(eqn.~(\ref{cor}). 
The process of maximization over the parameters is however a nonlinear
process and hence the statistics of $c(\mu^j)$ 
will be non-Gaussian. For a given
realization of noise, we assume that the global maximum of $c(\mu^j)$
will also be a local maximum. This assumption depends on the specific
nature of the waveform and the filter bank. The equations which 
will be satisfied at the estimated parameters 
$\mu^j={\hat\mu^j}$, are,
\begin{eqnarray}
\label{m1}
&&\frac{\partial c}{\partial\mu^i}  =0\mbox{\ \ \ }\nonumber\\
&\Rightarrow& 
\left\langle\check A {\bf h}(\check\mu^j) + {\bf n}, \frac{\partial
{\bf h}}{\partial\mu^i}\right\rangle = 0\nonumber\\
&\Rightarrow&  \left\langle{\bf h}(\check\mu^j), \frac{\partial
{\bf h}}{\partial\mu^i}\right\rangle = -\frac{1}{\check
A}\left\langle{\bf n}, \frac{\partial
{\bf h}}{\partial\mu^i}\right\rangle.
\end{eqnarray}

In the limit of high SNR the mean square errors in the measurement of the
parameters are characterised by the so called covariance matrix $C^{ab}$
\cite{HEL} defined as,
\begin{eqnarray}
\label{cov}
\Gamma_{ab} &=& \left\langle\frac{\partial{\bf s}}{\partial \mu^a}(\bbox{\mu}),
 \frac{\partial 
{\bf s}}{\partial \mu^b}(\bbox{\mu})\right\rangle\\
C^{ab} &=& \left[\Gamma^{-1}\right]^{ab}
\end{eqnarray}
   
From the expression for the Fourier transform in the stationary phase
approximation (eqn.~(\ref{FT})) and from the definition of the scalar
product (eqn.~(\ref{scal})) we see that the phase function in  the
Fourier transform simply cancel out and only the amplitude remains. We
also  find that $C^{ab} \propto \check A^{-2}$. 

The idea now is to compare the results of the Monte Carlo simulations
with those obtained via  eqn.~(\ref{cov}). In the limit of high SNR,
it is expected that the errors will agree with those obtained from the
covariance matrix. In the following sections we briefly mention the
results of the Monte Carlo simulations and then analyse in detail
eqn.~(\ref{m1}). The goal is to first check the agreement between
the two and then understand how the additional errors arise by
studying the consequences of eqn.~(\ref{m1}).

\section{The Newtonian Waveform}
\label{MCres}
\subsection{Monte Carlo Simulations}
\label{MCres1}
In this section we shall restrict ourselves to the Newtonian
waveform. We have the four parameters:
$$
\bbox{\mu} \equiv \mu^a \equiv \{\mu^0,\mu^1,\mu^2,\mu^3\} \equiv 
\{A, 2\pi f_st_s, \Phi, 2\pi
f_s\tau_0\}.
$$
Simulations were carried out for a test case of $\check\tau_0 = 5.558$
s This corresponds to a binary comprised of a $10M_\odot$ black
hole and a $1.4M_\odot$ neutron star. The waveform was cutoff at a
frequency of $800$ Hz and the data train was sampled at $2000$Hz. As
mentioned before the power spectral density was chosen to be
consistent with the initial LIGO interferometer. The range of $\tau_0$
for the filters was $[5.358,5.758]$ with a spacing of $1$ msec. 
The simulations presented here were carried out for an SNR of $10$, 
{\em i.e.}, $\check A = 10$. We considered $12000$ realizations of noise.

The covariance matrix for the Newtonian case is given below:
\begin{equation}
\label{cov1}
{\bf C}_{(\bbox{\mu})}\equiv 
\left[C^{ab}_{(\bbox{\mu})}\right] = \frac{1}{{\check A}^2}\left(\begin{tabular}{cccc}
${\check  A}^2$&0.0&0.0&0.0\\
0.0\ \ &222.50&322.26&-227.25\\
0.0&322.26&469.16&-328.65\\
0.0&-227.25&-328.65&232.26\\
\end{tabular}\right)
\end{equation}
In the above the diagonal values denote the variances in the
parameters and the nondiagonal components denote the covariances
between the parameters. Since the components do not depend on the
parameters $\mu^j$, (see \cite{Sat94,BSD96} for details)
we can choose a new set of parameters in which the covariance matrix
is diagonal. The transformation can be affected by means of an orthogonal
transformation. We have the new parameter set
$\bbox{\nu}\equiv\{A,\nu^1,\nu^2,\nu^3\}$,
related to $\bbox{\mu}$ by means of the relation $\bbox{\nu}={\bf S}
\bbox{\mu}$, where $\bf S$ is a orthogonal matrix, which for our
particular case is:
\begin{equation}
\label{smat}
{\bf S} = \left(\begin{tabular}{cccc}
$1.0$&0.0&0.0&0.0\\
0.0\ \ &-0.794  & 0.129 &  -0.594\\
0.0&0.359&  -0.690 & -0.629\\
0.0&-0.491 &  -0.713 & 0.501\\
\end{tabular}\right)
\end{equation}
In this new coordinate system the covariance matrix ${\bf C}_{(\nu)} =
{\bf S}{\bf C}_{(\mu)}{\bf S^{-1}}$, and is given by, 
\begin{equation}
 \label{cov2}
{\bf C}_{(\bbox{\nu})} = \frac{1}{{\check A}^2}\left(\begin{tabular}{cccc}
${\check  A}^2$&0.0&0.0&0.0\\
0.0\ \ &0.024&0.0&0.0\\
0.0&0.0&1.640&0.0\\
0.0&0.0&0.0&922.3\\
\end{tabular}\right)
\end{equation}
In the high SNR limit the root mean square errors in the parameters $\nu^a$
are given by,
\begin{equation}
\sigma_{\nu^a} = \sqrt{C^{aa}}.
\end{equation} 
For an SNR of $10$ the values are $\{1.0,\ 0.015,\ 0.128,\ 3.037\}$.

In a previous work \cite{BSD96} we had performed detailed  Monte Carlo
simulations to study the variation of the  errors with the SNR.
The simulations carried out in that paper were with a slightly
different power spectrum, as we have used a simple fit to
the noise curve in this paper.
We reproduce in Fig.~\ref{bsdfig}, the variation of
errors in the parameters with the 
SNR as given in Figure 5 of \cite{BSD96} for the
parameters $\tau_0$ and $t_a$, except that we
use the log scale for both the axis, as is conventional in statistical
literature. The continuous line represents the errors computed via the
covariance matrix and in this approximation the errors are inversely
propotional to the SNR. The dotted line represents the errors as
obtained from the Monte Carlo simulations. The rest of this paper 
tries to explain the discrepancy between the two curves 
in this figure at low SNRs $\rho\approx10$. 

\subsection{The equation for the errors}
\label{MCres2}

The expression for the Fourier transform of the chirp in
eqn.~(\ref{FT})
 in the new coordinates retains its form, but the functions
$\chi_i(f)$ are now transformed to,
$$
\eta_i(f) = \sum\limits_{j=1}^3\chi_j \left[S^{-1}\right]^j_i.
$$
Rewriting eqn.~(\ref{m1}) in the new parameters we have at $\nu^j=\hat\nu^j$,
\begin{equation}
\label{m1nu}
\kappa_i = -\frac{1}{\check
A}\left\langle{\bf n}, \frac{\partial
{\bf h}}{\partial\nu^i}\right\rangle
= \left\langle{\bf h}(\check\nu^j), \frac{\partial
{\bf h}}{\partial\nu^i}\right\rangle. 
\end{equation}
Using the definition of the scalar product in eqn.~(\ref{scal}), we
have, 
\begin{equation}
\label{kap2}
\kappa_i = 2 {\cal N}^2
\int_{0}^{\infty}\frac{f^{7/3}\eta_i(f)\sin(\sum\limits_{j=1}^3 
\eta_j(f)\Delta\nu^j)}{S_n(f)}df,
\end{equation}
where $\Delta\nu^j = \hat\nu^j-\check\nu^j$.
In the Newtonian case these constitute a set of three nonlinear equations
connecting $\nu^i$ to  $\kappa_i$. 
The quatities $\kappa_i$ are random variables. Since $\hat\nu^i$
depend upon $\bf n$, $\kappa_i$ are in general nonlinear functions of
the noise. This implies that the statistics of $\kappa_i$ can in
general be non-Gaussian. However, for  chirp signals of the type we
consider, the statistics of $\kappa_1$ and $\kappa_2$ will turn out to
be Gaussian. 
This will be demonstrated in the next section.
We define the total phase difference $\theta$ between the signal and
the filter as,
\begin{equation}
\label{def1}
\theta(f,\Delta\nu^i) = \sum\limits_{j=1}^3 
\eta_j(f)\Delta\nu^j,
\end{equation}
which is relevant for future considerations.
In the high SNR limit the errors
$\Delta\nu^j$ are small and hence we can make the approximation
$\sin(\theta)\simeq\theta$. This corresponds to using the covariance matrix
to provide an estimate of the errors. 
At astrophysically interesting SNRs $\approx 10$ this assumption does
not hold. 

It is clear that a good correlation between two waveforms is obtained
when the phase difference between the two waveforms is a multiple of
$2\pi$ and is roughly constant over the time for which the waveforms last.
This assumption is found to be true in the simulations. Even though
there are far more outlying points  than what is predicted by the
covariance matrix the phase difference between the signal and the
optimally matching template is found to be roughly constant. A typical
case is illustrated in Fig.~\ref{fig2}, in which $\theta$ is plotted
as a function of frequency for $\Delta\nu^i\equiv\{1.62, -8.49,
7.3\}$. The function $\theta(f)$ is found to be close to $-4\pi$ in
the region around $135$ Hz, from which the maximum contribution to the
SNR is obtained. Note that the values of $\Delta\nu^1$ and
$\Delta\nu^2$ are way beyond the rms errors predicted by the
covariance matrix.

Since the function $I(f)$ in Fig.~\ref{fig1} peaks at
$f\approx135$Hz.,
we define $\theta_m$ to be the value of $\theta$ at $f=135$Hz., 
{\em i.e.},
\begin{equation}
\label{thetam}
\theta_m = -3.87615\Delta\nu^1 + 0.739347\Delta\nu^2 + -0.0149334\Delta\nu^3.
\end{equation}
In Fig.~\ref{fig3} we illustrate how the  $\theta_m$ is distributed
 across many realizations of noise. It is clear from the figure 
that $\theta_m$ is strongly clustered around multiples of $2\pi$.
Since $\theta_m$ is a linear combination of the parameters, it is
clear that at least some of the parameters will show similar
clustering properties.  The variable $\theta_m$ is a very convenient
indicator of the rough location of the parameters in the parameter
space. 
If the absolute value of $\theta_m$ is large then the
estimated parameters are far away from the actual value of the
parameters.

Fig.~\ref{fig4} is a $\nu^1-\nu^2$ scatter plot, where each 
point corresponds to a realization of noise. The plot shows that
$\nu^1$ and $\nu^2$ parameters are clustered in well separated 
`islands' in the parameter space considered.
 The variable $\theta_m$ computed for points in any specific 
island  yield values close to a specific integral multiple of $2\pi$
depending upon the island we choose.
Multiple solutions to eqn.~(\ref{kap2}) are responsible for the
islands. Since the pdf in the $\kappa_i$ space is
largest at the origin, the islands correspond to $\theta(f)\simeq
2k\pi$,  where $k$ is an integer. 
We give an explanation of this below. The phase parameter
$\mu^2\equiv\Phi$ is constrained to lie in the interval $[-\pi,\pi]$.
Using the relation $\mu^i=\left[S^{-1}\right]^i_j\nu^j$, we have,
\begin{equation}
\label{phase1}
\Phi\equiv\mu^2=0.129212\nu^1 + -0.689524\nu^2 + -0.712644\nu^3.
\end{equation}
Since the rms error in $\nu^3$ as calculated from the covariance
matrix at an SNR of $10$ is $3.04$, it is highly likely that the 
absolute value of the last
term in eqn.~( \ref{phase1}), ( which dominates the other terms in
the same equation, since the rms errors for the other parameters are
much lesser)  exceeds $\pi$. This forces the
parameters $\nu^1$ and $\nu^2$ to {\em jump} to values such that 
$\Phi$ remains in the range $[-\pi,\pi]$. In order to calculate 
the amount by which the parameters jump, we consider
eqn.~(\ref{thetam}). 
Since $\theta_m$ must be a multiple of $2\pi$ in order
to get a good match, each of the first two terms on the right hand
side of eqn.~(\ref{thetam}) contributes $\pi$. This implies that $\nu^1$
jumps by $\pi/3.87615 \approx .81$ and  $\nu^2$
jumps by $\pi/0.739347 \approx 4.25$. Since the two terms are of
opposite signs, we find that $\nu^1$ increases when $\nu^2$ 
decreases and vice versa. 

We now revert back to eqn.~(\ref{kap2}). Since $\theta$ is close to
an integral multiple of $2\pi$ in the frequency region of interest, we
can write,
\begin{equation}
\label{approx}
\sin(\theta) \approx \left(\sum\limits_{j=1}^3\eta_j(f)\Delta\nu^j - 2k\pi\right),
\end{equation} 
for some integer $k$.  We have,
\begin{equation}
\label{kap3}
\kappa_i = 2 {\cal
N}^2\sum\limits_{j=1}^3\int_{0}^{\infty}\frac{f^{7/3}
\eta_i(f)\eta_j(f)\Delta\nu^j}{S_n(f)}df - 4k\pi{\cal N}^2\int_{0}^{\infty}\frac{f^{7/3}
\eta_i(f)}{S_n(f)}df. 
\end{equation}
Since the $\bf\nu$ coordinate system is orthogonal, 
\begin{equation}
\label{ortho}
\int_{0}^{\infty}\frac{f^{7/3}
\eta_i(f)\eta_j(f)\Delta\nu^j}{S_n(f)}df = 0 \ \ \ \mbox{for\ \ }
i\neq j,
\end{equation}
and
\begin{equation}
\label{ortho2}
2 {\cal N}^2\int_{0}^{\infty}\frac{f^{7/3}
\eta_i(f)^2}{S_n(f)}df = \frac{1}{C^{ii}_{(\nu)}}.
\end{equation}
Therefore,
\begin{equation}
\label{kap4}
\kappa_i= \frac{\Delta\nu^i}{C^{ii}_{\nu}} - 2k\pi G_i,
\end{equation}
where,
\begin{equation}
\label{kap5}
G_i = 2 {\cal N}^2\int_{0}^{\infty}\frac{f^{7/3}
\eta_i(f)}{S_n(f)}df.
\end{equation}
The amount  by which the parameters $\nu^1$ and
$\nu^2$ jump can be calculated more accurately using 
eqn.~(\ref{kap4}). For two successive values of the integer $k$, the 
parameter $\nu^1$ has to jump by, $2\pi G_1/C^{11}_{\nu}$ in order
to have the same value of $\kappa_1$. This works out to be
$0.812$ for the parameter $\nu^1$. Similarly the value for 
the parameter $\nu^2$ turns out to be $4.332$. 
These values are  consistent with the scatter plot in Fig.~\ref{fig4}.

It is clear from Fig.~\ref{fig4} that the distribution of the 
parameters $\nu^1$ and $\nu^2$ will be markedly multimodal. However, 
the distribution of $\nu^3$ is relatively smoother. This is
illustrated in 
Fig.~\ref{fig5} which is  a scatter plot on the $\nu^2-\nu^3$ plane.
Though there are gaps along the $\nu^2$ axis, $\nu^3$ takes all values
in the range shown. The parameter $\nu_3$ is relatively well behaved,
{\em i.e.} it does not exhibit any sudden jumps like the other
parameters.

The variances as obtained from the Monte Carlo simulations in the
parameters $\nu^i$ are, $\Sigma_{\nu^i}\equiv\{0.421, 2.26, 2.53\}$ whereas,
the values  predicted by the
covariance matrix are $\sigma_{\nu^i}\equiv
\{0.0153, 0.1281, 3.037\}$. In the  case of the parameters
$\nu^1$ and $\nu^2$ the observed variances are much larger than the 
lower bounds, due to the jumps which these parameters make. However we
notice that the observed variance in the case of $\nu^3$ is actually
{\em less} than the lower bound. This is due to the fact that the
Cramer-Rao bounds are applicable only to parameters which are allowed
to vary freely. For instance the variance for the parameter
$\Phi\equiv\mu^2$ can have the maximum value of $\pi^2$ whatever be
the SNR. The restriction of the range of $\mu^2$ puts a constraint on
the values of the parameters $\nu^j$ and this accounts for the
observed error being less than the Cramer-Rao bound. 
In the next section we give a more quantitative model for
the distribution of the parameters.	

\subsection{Geometrical Perspective}
\label{geom}
In this section we use differential geometry to arrive at a
statistical model for the distribution of the parameters.
The model described here is quite general and is applicable  to the
estimation of parameters obtained by means of the maximum likelihood  method,
for an arbitrary signal in the presence of
Gaussian noise. 
The set of signal vectors, ${\bf s}(\bbox{\nu}) \equiv
s(t;\bbox{\nu})$, where $\bbox{\nu} 
\equiv \{\nu_0,\ldots,\nu_{m-1}\}$, is a m-dimensional parameter vector,
will describe a m-dimensional manifold $\cal S$ embedded in ${\cal
V}$, the vector space of all detector outputs. (See
\cite{BSD96,Ow96} for an introduction to the use of differential
geometry in gravitational wave data analysis, and \cite{Am} for the
application of differential geometry to statistics.) 
Let the output of a detector $\bf x$, contain a signal 
${\bf s}(\bbox{\check{\nu}})$. 
Then  $\bf x = {\bf s}(\bbox{\check{\nu}})  + \bf n$,
where $\bf n$ is a noise vector drawn from the noise distribution. 
The distance, $D(\bbox{\nu})$ between  $\bf x$ and a point 
$ {\bf s}(\bbox{\nu})$ is given by, 
\begin{eqnarray}
\label{dist}
D(\bbox{\nu}) &=& \| \bf{x - s}(\bbox{\nu})\| = 
\left\langle\bf{x - s}(\bbox{\nu}), {\bf x -
s}(\bbox{\nu})\right\rangle^{1/2},\\
&=& \left[\left\langle{\bf x},{\bf x}\right\rangle - 
2\left\langle{\bf x},\bf{s}(\bbox{\nu})\right\rangle + 
\left\langle {\bf s}(\bbox{\nu}),
{\bf s}(\bbox{\nu})\right\rangle\right]^{1/2}. 
\end{eqnarray}
 The MLE of the parameters is that
point $\bbox{\hat{\nu}}$ on the parameter space which minimises
$D(\bbox{\nu})$. This is equivalent to
maximising the correlation 
$c(\bbox{\nu}) = \left\langle {\bf x},{\bf s}(\bbox{\nu})
\right\rangle$ provided we keep $\left\langle {\bf s}(\bbox{\nu}),
{\bf s}(\bbox{\nu})\right\rangle$ constant.

In the limit of high SNR, ${\bf s}(\bbox{\hat{\nu}})$ will lie
in a small
region around ${\bf s}(\bbox{\check{\nu}})$  on the manifold,
effectively the tangent space to the manifold at
${\bf s}(\bbox{\check{\nu}})$. 
In this case, the difference, ${\bf s}(\bbox{\hat{\nu}}) -
 {\bf s}(\bbox{\check{\nu}})$ can be satisfactorily approximated as 
the projection of the noise vector onto the tangent space. This
implies that ${\bf s}(\bbox{\hat{\nu}}) -
 {\bf s}(\bbox{\check{\nu}})$ is 
linear function of $\bf n$. 
Further, if the parameters form a Cartesian system of coordinates, 
then, they too will be linear in $\bf n$ and the distribution of the 
parameters can be described by a multivariate Gaussian \cite{Fin92}. 
The covariance matrix of this distribution defines a lower bound on the
errors in estimation and is termed as the Cramer-Rao bound. 

If the global minimum of $D(\bbox{\nu})$ is also a local minimum then,
at $\bbox{\nu}=\bbox{\hat\nu}$,
\begin{equation}
\partial D(\bbox{\nu})/\partial\hat\nu^a = 0,\ \ \mbox{for}\ \ a=0,\ldots,m-1
\end{equation}
which implies,
\begin{equation}
\label{max}
\left\langle{\bf s}(\bbox{\check{\nu}}) + {\bf n} 
  - {\bf s}(\bbox{\hat{\nu}})
 ,\frac{\partial 
{\bf s}}{\partial \nu^a}(\bbox{\hat{\nu}})\right\rangle = 0. 
\end{equation}
These are a set of $m$ equations, one for each parameter.
We interpret these equations geometrically as follows.
The equations imply that the vector 
${\bf x} -  {\bf s}(\bbox{\hat{\nu}})$ is orthogonal to each of the coordinate
basis vectors, $\partial/\partial\nu^a$, evaluated at 
$\bbox{{\nu}}=\bbox{\hat{\nu}}$.  
Thus $\bbox{\hat{\nu}}$ is a local extremum when the tip of the vector $\bf x$
lies on that $N-m$ dimensional hyperplane ${\cal B}_{\bbox{\hat{\nu}}}$, 
which passes through  
${\bf s}(\bbox{\hat{\nu}})$, and is orthogonal to the $m$-dimensional 
tangent plane at ${\bf s}(\bbox{\hat{\nu}})$.
This hyperplane, ${\cal B}_{\bbox{\hat{\nu}}}$,  is the
intersection of the $(N-1)$-dimensional hyperplanes, 
${\cal N}^a_{\bbox{\hat{\nu}}}$, each orthogonal
to a coordinate basis vector ${\partial}/{\partial\nu^a}$ at 
$\bbox{{\nu}}=\bbox{\hat{\nu}}$.

Let ${\bf l}_a$ be the normalized coordinate basis vectors at
$\bbox{\hat{\nu}}$, {\em i.e.}
\begin{equation}
\label{ells}
{\bf l}_a = \frac{\partial {\bf s}}{\partial \nu^a}(\bbox{\hat{\nu}})\bigg/
\left\| \frac{\partial {\bf s}}{\partial \nu^a}(\bbox{\hat{\nu}})\right\|.
\end{equation}
We define 
$r_a$ to be the minimal distance from ${\bf s}(\bbox{\check{\nu}})$ to the
hyperplane ${\cal N}^a_{\bbox{\hat{\nu}}}$, which is given by 
\begin{equation}
r_a=\left\langle{\bf s}(\bbox{\hat{\nu}})- {\bf s}(\bbox{\check{\nu}})
,{\bf l}_a\right\rangle.
\end{equation}
A schematic illustration of the above discussion is given in 
Fig.~\ref{fig6}.

The pdf for the vector $\bf x$ to lie on 
${\cal N}^a_{\bbox{\hat{\nu}}}$ can  depend only on ${\bf l}_a$ and
$r_a$. If the  vector $\bf x$ is to lie on 
${\cal B}_{\bbox{\hat{\nu}}}$, then it must simultaneously lie on each
of the normal hyperplanes, ${\cal N}^a_{\bbox{\hat{\nu}}}$.  The
pdf  for  $\bf x$  to lie on 
${\cal B}_{\bbox{\hat{\nu}}}$ is given by the expression,
\begin{equation}
\label{rdis2}
p(r_a) = \int_{\cal V}\left[\prod_{a=0}^
{m-1}\delta(\left\langle 
({\bf n} - r_a{\bf l}_a),{\bf l}_a\right\rangle)\right] \ p({\bf n})
\ d^{\scriptscriptstyle N}n,                  
\end{equation}
where the $\delta$ denotes the Dirac Delta function. 
Substituting for the Gaussian distribution for the noise
$p({\bf n})$ in the equation above and integrating,
we get,  
\begin{equation}
\label{rdis1}
p(r_0,r_2,\ldots,r_{m-1}) = {{\exp\left[
-\frac{1}{2}\sum\limits_{a,b=0}^{m-1}\left[\gamma^{-1}\right]^{ab}
r_ar_b\right]}\over{\left[\left(2\pi\right)^m
\mbox{det}\left[\gamma_{ab}\right]\right]^{1/2}}} ,
\end{equation}
where, $\gamma_{ab} = \left\langle{\bf l}_a,{\bf
l}_b\right\rangle$. The integration though tedious, is quite
straightforward. Note that  each of the Gaussian
random variables $r_a$ will have unit variance as is obvious from the
definition of $\gamma_{ab}$. Moreover, the matrix 
$\gamma_{ab}$ is very closely
related to the Fisher information matrix $\Gamma_{ab}$ as defined in
eqn.~(\ref{cov}). Whereas the components of the Fisher information
are got by taking scalar products of the tangent vectors on the
manifold, the components of the matrix $\gamma_{ab}$ are obtained by
by taking scalar products of the {\em normalized} tangent vectors on the
manifold.

\subsection{Statistical model for the Newtonian
Chirp}
\label{appnewt}
We now specialise to the case of the Newtonian chirp. We use the
parameters $\nu^a$ defined earlier, in the previous section.
Since  ${\bf s}(\bbox{\hat{\nu}}) = \hat A {\bf h}(\hat\nu^j)$,
the above equations for ${\bf l}_a$  and $r_a$ give,
\begin{eqnarray}
{\bf l}_0 &=& {\bf h}(\hat{\nu}^j),\\
{\bf l}_k &=& \frac{\partial {\bf
h}}{\partial\nu^k}\left(\hat{\nu}^j\right)\bigg/\left\|
\frac{\partial {\bf h}}{\partial\nu^k}\left(\hat{\nu}^j\right)
\right\|,\\
r_0(\hat\nu^k,\hat A) &=& \check A \left\langle{\bf h}(\check{\nu}^k),{\bf h}(\hat{\nu}^k)
\right\rangle - \hat A,\\
\label{r2}
r_j(\hat\nu^k) &=&  \check A\left\langle {\bf h}(\check\nu^k),\frac{\partial {\bf
h}}{\partial\nu^j}\left(\hat{\nu}^k\right)\right\rangle\bigg/\left\|
\frac{\partial {\bf h}}{\partial\nu^j}\left(\hat{\nu}^k\right)
\right\|.
\end{eqnarray}
Since the $\bbox{\nu}$ coordinate system is orthogonal, $\gamma_{ab}$ turns
out to be nothing but the identity matrix. 
The
distribution of the variables $r_a$ is thus a joint Gaussian given by the 
expression,
\begin{equation}
\label{rdisnu}
p(r_0,\ldots,r_3) = \frac{1}{2\pi^2}\exp\left[
-\frac{1}{2}\sum\limits_{a=0}^{3}r_a^2\right].
\end{equation}

The $r_j$ variables are closely related to the $\kappa_i$ variables
defined in section \ref{MCres2} in eqn.~(\ref{m1nu}):
$$
r_i \ =\  \check A \left\|
\frac{\partial {\bf h}}{\partial\nu^i}\left(\hat{\nu}^k\right)
\right\|^{-1}\kappa_i  
\ =\  \frac{\kappa_i}{\sqrt{\Gamma_{ii}}}, 
$$ 
where we have used the definition of $\Gamma_{ab}$ in eqn.~(\ref{cov}).
Thus they differ only by a factor which is  
simply a constant from eqs.~(\ref{scal}) and (\ref{FT}). This is a
consequence of the intrinsic flatness of the manifold \cite{BSD96} and
the particular parameterization adopted.

From eqn.~(\ref{rdisnu}) we would expect the marginal probability distribution of 
each of the variables $r_j$ to be a Gaussian distribution with unit
variance. Using the ensemble of estimated parameters from the Monte
Carlo simulations we can obtain the histograms and consequently the
distributions of the variables $r_j$.   
In Fig.~\ref{fig7} we plot the probability distributions of the 
variables $r_j$, and compare them with the expected Gaussian distributions.
It is clear that though $r_1$ and $r_2$ are Gaussian random variables
to a good approximation, $r_3$ shows a pronounced non-Gaussian behaviour. 
The reason for this discrepancy can be traced to the fact that the
phase parameter is constrained to lie in the range $[-\pi,\pi]$. 
In Fig.~\ref{fig5} we observe that in the central island the
$\nu^3$ parameter gets abruptly cutoff at a value of about $4.5$
and $-4.5$. Since the points on the central island are `close' to 
the point corresponding to the actual value of the parameters, we can
apply the eqn.~(\ref{kap4}) with $k=0$. This gives a value of 
$\kappa_3\approx 0.00488$ and consequently $r_3\approx1.5$.
We
observe in Fig.~\ref{fig7} that the dip in the distribution of 
$r_3$ occurs at the same point {\em i.e.} at $r_3\approx1.5$.
To further elaborate on this point, we plot in the Fig.~\ref{fig8},
the three variables $r_i$ v/s the variable $\theta_m$ defined earlier.
The figure clearly illustrates that while $r_1$ and $r_2$
take on there entire range of values in the central island, $r_3$ does
not. 
This establishes a connection between
the dip in the marginal distribution and the phase parameter being
constrained to the range $[-\pi,\pi]$.
Although a deeper understanding of this is in order,
we continue our analysis assuming the distribution of 
$r_3$ to be given by a Gaussian.

Had the map between $\bbox{\hat{\nu}}$ to $\bf r$ been bijective,
it would have been possible to write the distribution for the
estimated parameters simply as a product of the pdf for the variables
$r_a$ and a Jacobian factor. However,
a given set of values for ${\bf r} \equiv 
\{r_a\}$ would in general 
correspond to multiple  parameter vectors $\bbox{\hat{\nu}}^{(l)}$,
where the range of values of $l$ depends on the number of solutions. 
This is clear from Fig.~\ref{fig8} where we observe that the same value
of $r_j$ occurs for different values of the variable $\theta_m$,
and hence in different `islands'.
We also observe that
$r_3$ takes almost all its possible values in the central island and the two
adjoining ones. Moreover, we notice that it is approximately true that
$r_3$ takes different values in each of the three islands. Thus if we restrict
ourselves only to the central and two adjoining islands, then the map
between $\nu^j$ and $r_j$  is bijective to a good degree of
approximation. For a fixed set of values for $\{r_j\}$
 we have therefore, a unique solution $\nu^j$ satisfying
eqs.~(\ref{r2}), if we restrict ourselves to 
 the three islands identified above. We shall henceforth term 
the islands identified above as
the {\em primary group} of islands and the 
solution  there will be called the {\em primary} solution. 
(It is to be emphasized that the above discussion is applicable only to
the Newtonian waveform. As we shall see later for the post-Newtonian
case, the primary group of islands contains more islands on either
side of the central island.)
There will of course be other solutions in  the other islands.
It is to be noted that number of multiple solutions for a 
given set of values for $\{r_j\}$, depends on $\{r_j\}$, and we do
{\em not} imply that each island admits a solution $\nu^j$ for a fixed
 set of values for $\{r_j\}$.
The reason for the term {\em primary} solution is to be found in 
Fig.~\ref{fig9}, where we have plotted a histogram of the variable
$\theta_m$ which gives us an idea as to how many points occur in each
island. We observe that  $99\%$ of the points lie in the central and
the two adjoining islands for the Newtonian case.
 Thus there is an overwhelming probability
for the points to lie in one of these islands, and the primary
solutions occur much more frequently as compared to the other
solutions for a fixed value of $r_j$.
 
The problem now is to determine the various solutions
$\bbox{\hat{\nu}}^{(l)}$
for a fixed set of values of  $\{r_a\}$, and
compute  the probability that a particular
$\bbox{\hat{\nu}}^{(m)}$ will be selected amongst others 
for a  given $\bf r$.  This  is essentially the probability that the
amplitude $\hat A^{(m)}$ is greater than the amplitude at the other
solutions. We shall denote this probability as 
$P(\bbox{\hat{\nu}}^{(m)}|{\bf r})$. 
For a  fixed set of values of  $\{r_j\}$, we will have multiple
solutions, $\{\hat\nu^j\}^{(1)}, \{\hat\nu^j\}^{(2)},\ldots$\ 
to eqn. (\ref{r2}).
The correlation obtained at these points 
will be 
$$
\hat A^{(l)} = \left\langle{\bf
x},{\bf h}(\hat\nu^j)^{(l)}\right\rangle.
$$
It is to be noted that
for a {\em fixed}
$\{\hat\nu^j\}$, $\hat A$ will be a Gaussian random variable with a
variance of unity.
In order to calculate 
$P(\{\hat\nu^j\}^{(l)}|\{r_j\})$ we need to identify all the solutions
corresponding to a fixed set of values $r_a$.
The identification of the multiple roots is quite a problem, and so we
make the following approximation. We assume that for one of the
solutions, that is the primary solution which we 
shall denote by 
$\{\hat\nu^j\}^{(1)}$, corresponding to $\{r_j\}$, the probability 
$P(\{\hat\nu^j\}^{(1)}|\{r_j\})$ is almost unity. 
If this is
true, then, we only need to compute the probability that the correlation
at an arbitrary point on the parameter space exceeds the correlation
at the primary solution point which shares the same set of values  of $\{r_j\}$.
This corresponds to computing the probability of the Gaussian random
variable $\hat A^{(l)}$, exceeding $\hat A^{(1)}$. 
Of course, $P(\{\hat\nu^j\}^{(1)}|\{r_j\})$ is set to unity.
The justification for the above procedure is essentially the fact that
nearly $99\%$ of the points lie in the primary group of islands.

So for evaluating the distribution of the estimated parameters at 
 $\nu^k=\hat\nu^k$ we follow the following procedure:
\begin{enumerate}
\item  	Determine $r_j(\hat\nu^k)$ using
	eqn. (\ref{r2}).
\item  	Determine $\{\hat\nu^k\}^{(1)}$ using eqn.
	(\ref{kap4}) for an appropriate value of $k$,	
	{\em i.e.} $k$ takes one of the values $\{-1,0,1\}$.
\item 	Determine the probablility for $\hat A 
	= \left\langle{\bf x},{\bf h}(\hat\nu^k)\right\rangle$ to be  greater than 
	$\hat A^{(l)} = \left\langle{\bf
	x},{\bf h}(\hat\nu^k)^{(1)}\right\rangle$. (If $\hat\nu^k$ is
        already a primary solution then this probability is set to unity.)
\item   Set  $P(\{\nu^a\}|\{r_a\})$ equal to the calculated
	probability.  
\item	Write the the pdf of the estimated parameters as  
\begin{equation}
\label{ndis1}
p(\bbox{\hat\nu})= 
\frac{1}{\left(2\pi\right)^2} 	
J\left(\bbox{\hat\nu}\right)P\left(\bbox{\hat\nu}|{\bf r}\right)
\exp\left[-\frac{1}{2}\sum\limits_{a=0}^3
r_a^2(\bbox{\hat\nu})\right],
\end{equation}
where, $ J(\bbox{\hat{\nu}})$
is the Jacobian of the transformation from ${\hat{\nu}^a}$ to the
variables $r_a$, which is essentially,
\begin{equation}
\label{jac}
J(\bbox{\hat\nu})=\mbox{det}\left[\frac{\partial r_a}{\partial\nu^b}(\bbox{\hat\nu})\right].
\end{equation}
\end{enumerate}

Since the amplitude parameter
is not of primary interest to us, we shall integrate the distribution
over $\hat A$. We use a threshhold of $7$, which means that we reject
any realization whose measured value $\hat A$ is less than $7$ in the
simulations. The amplitude parameter enters only via $r_0$. So the
distribution for the remaining three parameters $\hat\nu^k$ is,
\begin{eqnarray}
\label{ndis2}
p(&&\hat{\nu^k})= 
\frac{1}{\left(2\pi\right)^{3/2}} 	
J\left(\hat\nu^k\right)P\left(\hat\nu^k|{\bf r}\right)
\exp\left[-\frac{1}{2}\sum\limits_{i=1}^3 r_i^2(\hat\nu^k)\right]
\;\mbox{\bf\LARGE$\times$}\nonumber\\
&& \frac{1}{\left(2\pi\right)^{1/2}}\int_{7.0}^\infty
\exp\left[-\frac{1}{2}
\left(\check{A}\left\langle h({\check{\nu^k}}),h({\hat{\nu^k}})
	\right\rangle-\hat{A}\right)^2\right] d\hat{A}.
\end{eqnarray}
Since the SNR is chosen to be 10 the second factor in the 
equation above is very close to unity.

We now compare the Monte Carlo results with the distribution given in
eqn.~(\ref{ndis2}). We  compare the one dimensional marginal 
distribution $p(\hat\nu^1)$
with the histogram obtained via the Monte Carlo method. 
The marginal 
distribution $p(\hat\nu^1)$ is obtained by integrating 
eqn.~(\ref{ndis2}) with respect to $\nu^2$ and $\nu^3$. This is done
numerically.
Though the parameter $\nu^1$ has the
least root mean square error of $.015$ 
as predicted by the covariance matrix, its distribution
has the most pronounced non-Gaussian behaviour. In plot (a) and (b) of
Fig.~\ref{fig10} we display the distributions $p(\hat\nu^1)$,
obtained from the Monte Carlo simulations and the statistical model 
respectively. 
Plots (c) and (d) in the same figure zoom in on the 
first two maxima occurring on the right of  the central maximum. 
It can be seen that in the case of plot (d) the match is 
not very good even though
the location of the peaks match fairly well. 
The difference here could come from
either the Monte Carlo method or the statistical model. 

In the histogram of the variable $\theta_m$ illustrated in 
Fig.~\ref{fig9} we have seen that about $79.5\%$ of the points in the
simulations fall in the central island and about $10\%$ in each of the    
adjoining islands at an SNR of $10$.  We can obtain the corresponding
numbers from our theoretical model as given in eqn.~(\ref{ndis1}), by
integrating the distribution over all the parameters in the  domain
corresponding to each island. We obtain  the values  
$82\%$ for the central island  and about $9.5\%$ for each of the
adjoining islands. Thus the statistical model shows good agreement
with the Monte Carlo results by this criterion.
The number of points in the
subsequent islands falls off rapidly. The contribution to the variance
from each island depends on the number of points in that region and
the location of island in the parameter space. It is found that the
maximum contribution to the variance comes from the islands
immediately adjoining the central island. 

\section{Post Newtonian waveform}
\label{PNwave}

In the post-Newtonian case we have the five parameters,
$$
\bbox{\mu} \equiv \mu^a \equiv \{\mu^0,\mu^1,\mu^2,\mu^3\} \equiv 
\{A, 2\pi f_st_s, \Phi, 2\pi f_s\tau_0, 2\pi f_s\tau_1\}.
$$
Simulations (12000  realizations) were carried out again for an SNR of
10. The signal parameters were $\check\tau_0 = 5.558$s\ and $\check\tau_1 =
0.684$s, corresponding to a binary comprised of a $10M_\odot$ black
hole and a $1.4M_\odot$ neutron star. The simulation box taken was
$\{5.058$s$\leq\tau_0\leq5.859$s$, 0.484$s$\leq\tau_1\leq0.985$s$\}$, with filter
spacings of $10$ms.\ in $\tau_0$ and $5$ms.\ in $\tau_1$.
Our analysis of the post-Newtonian case will be very similar to the
Newtonian case in section \ref{MCres}. All the variables defined there can
be defined for the post-Newtonian case also and we will use the same
names for the variables to avoid using more symbols.

Here again we diagonalise the covariance matrix by making a coordinate
transformation from the $\bbox{\mu}$ to the $\bbox{\nu}$ coordinate
system. The diagonal covariance matrix in the  $\bbox{\nu}$ parameters
is given by,
$$
\left[C_{(\nu)}^{ab}\right] = \frac{1}{{\check A}^2}\left(\begin{tabular}{ccccc}
${\check  A}^2$&0.0&0.0&0.0&0.0\\
0.0\ \         &0.018&0.0&0.0&0.0\\
0.0            &0.0&1.1974&0.0&0.0\\
0.0            &0.0&0.0&403.64&0.0\\
0.0            &0.0&0.0&0.0&15601.0\\
\end{tabular}\right)
$$
Therefore the root mean square errors in the parameters computed from the covariance matrix
are,
$\sigma_{\nu^i}~\equiv~\{.013,0.109,2.009,12.49\}$. On the other hand 
the observed
errors from the Monte Carlo simulations  are $\Sigma_{\nu^i}~\equiv~\{1.33574, 7.43948,
5.34089, 23.049\}$. One finds as compared to the Newtonian case, the
factors
$\Sigma_{\nu^i}/\sigma_{\nu^i}$ are larger on the average. The reason
for this is that here we have one more parameter, which means there
are more filters which can match with the data and consequently there
are more outlying points. This is evident from Fig.~\ref{fig13} as is
explained below. 

In Figs.~\ref{fig11} and \ref{fig12} we give the scatter plots of
the four parameters $\nu^i$, the former on the $\nu^1-\nu^2$ plane
and the latter on the $\nu^3-\nu^4$ plane. We observe that while there
are gaps in distribution of the parameters $\nu^1$ and $\nu^2$ there
are none in the distribution of the parameters $\nu^3$ and $\nu^4$.
Here the phase parameter $\mu^2\equiv\Phi$ is given by,
$$
\Phi =  -0.109838\nu^1  - 0.621516\nu^2 + 0.695335\nu^3 +
0.343747\nu^4,
$$ and the variable $\theta_m$ is given by,
$$
\theta_m = 4.18241\Delta\nu^1+ 0.892242\Delta\nu^2 + 
0.0186845\Delta\nu^3 +  0.00272864\Delta\nu^4.
$$  
Since both $\sigma_{\nu^3}$ and $\sigma_{\nu^4}$ are comparable, both
these parameters will contribute to $\Phi$ as opposed to the Newtonian
case where only the $\nu^3$ parameter dominates. Thus we find that
both the  parameters $\nu^3$ and $\nu^4$ assume all their  values in
their respective ranges. The spacing between the
islands in the $\nu^1$ and $\nu^2$ scatter plot is calculated to be 
$2\pi G_1/C^{11}_{\nu} = 0.69$ and  $2\pi G_2/C^{22}_{\nu} =
3.90$  respectively. This is consistent with Fig.~\ref{fig11}.
In Fig.~ \ref{fig11} we notice that there are more islands to the
right than there are to the left of the central island. This is caused
by the simulation box which is asymmetrical, {\em i.e.} the parameters
of the signal are not at the center of the simulation box. This is so
since all combinations of $\tau_0$ and $\tau_1$ do not correspond to
valid masses for the components of the binary.  This leads to a shift
in the mean of the estimated parameters from their actual values.
The observed means in the  $\Delta\nu^i$ are  $\{0.10167, 0.574784, 
-0.282783, 1.56494\}$. For higher SNRs the estimated parameters will
tend to lie only on the islands which lie close to the central island
and consequently within a symmetrical region around the actual values
of the parameters. Therefore the bias will disappear for the case  of
higher SNRs.

In Fig.~\ref{fig13} we plot the histogram of the distribution of
the variable $\theta_m$ in the post Newtonian case. Here, for the same
SNR of $10$, there are more outlying points as compared to the
Newtonian case. This is to be expected since we have an additional
parameter, and there is an additional degree  of freedom to find the
filter which matches best with the signal.  The variance now gets
  substantial  contributions
even from far away islands as opposed to the Newtonian case. 

In Fig.~\ref{fig14} we plot the variables $r_i$ v/s $\theta_m$. 
Here again only $r_4$ does not attain all its possible values within
the central island, but now $r_4$ attains its entire range of values
only when when we include two islands on either side. Moreover the
overlap of values of $r_4$ in these islands is much more pronounced 
as opposed to the Newtonian case and further investigations are needed to
identify the primary solution correctly. 

In Fig.~\ref{fig15} we plot
the probability  distributions of the variables $r_i$. The histograms
are plotted using the Monte Carlo data and the continuous curve is a
Gaussian with a variance of unity. It is clear that $r_i$ are Gaussian
random variables to a good degree of approximation. The approximation
is better than in the Newtonian case (see Fig.~\ref{fig7}).
As in the
Newtonian case the $r_i$ and $\kappa_i$ are related by only a constant
factor. Therefore $\kappa_i$ are
also Gaussian random variables. However, we note that this is true
only for a special class of waveforms to which the chirp belongs. The
chirp signal manifold is intrinsically flat and the parameterization
of the waveform is such that the coordinates are essentially
Cartesian. If some other parameterization is chosen ({\em e.g.} the
chirp mass $\cal M$), the coordinates will no longer be so and
$\kappa_i$ will not be Gaussian random variables even though the
variables $r_i$ will remain Gaussian. 

We can again as in the Newtonian case, write down the expression for
the pdf of the estimated parameters. We do not write it explicitly
here since the expression is formally the same as in
eqn.~(\ref{ndis1}), except now the index $a$ runs from $0$ to $4$.
We therefore only refer to eqn.~(\ref{ndis1}) for the required
pdf.

\section{Conclusions}
\label{sec_con}
In this paper we have addressed the question of wide discrepancies between
the theoretically predicted lower bounds in the errors in the
estimation of parameters and the errors observed in numerical
experiments.
We now summarize in this section, the main results of our paper and
indicate future directions.
\begin{itemize}

\item  We find
that the problem is of a {\em global} nature, in that the estimated values
for the parameters fall into {\em disjoint} islands. 
Though there are very few
points in the islands which are far from the actual value of the
parameters, they contribute substantially to the variance.
Thus the discrepancy
between the Monte Carlo simulations and the  Cramer Rao bounds cannot
be resolved by using perturbative analysis.  The restriction of the parameter
$\Phi$ to the range $[-\pi,\pi]$  plays a major role in the
non local distribution of parameters, as explained in the text.

\item  The problem is more transparent when we reparameterize
the chirp signal so that the covariance matrix is diagonal in the new
parameters. The parameters $\bbox{\nu}$ correspond to choosing orthogonal
coordinates on the chirp manifold. Since the covariances are zero for
the variables $r_a$ the pdf is  computationally simpler to handle.

\item  A statistical model has been given which
matches well with the distribution obtained from Monte Carlo simulations.
The model is derived from geometrical considerations. We have
identified certain variables $r_i$ as Gaussian random variables and these
play an important role in writing down the expression for the
distribution of the estimated parameters. 

\item 
Since the distribution of the estimated parameters is multimodal,
the variance is not a good indicator of the performance of the MLE.
A more  reasonable indicator
would be to 
judge the performance of the MLE by means of confidence
intervals {\em i.e.} compute the probability that the estimated parameters lie
within a certain region around the actual value. 
As a concrete example, we compute the probability that 
the deviation of the estimated parameter is less than thrice  the root
mean square error at that SNR as predicted by the covariance matrix.
We will use the symbol $P_{3\sigma}$ to denote the fraction of points
which lie within the $3\sigma$ range for a given parameter. 
However this criterion will in general be dependent on the specific
parameter chosen. 
Since $\tau_0$ is one such physical parameter we use this to compute
$P_{3\sigma}$.

In Figures \ref{fig16} and \ref{fig17} we plot $P_{3\sigma}$ v/s SNR  for the
parameter $\tau_0$ in the Newtonian and the post-Newtonian cases respectively.
The values
of $P_{3\sigma}$ for the parameters such as $\tau_0$ and $\tau_1$ will
be independent of the actual value of the chirp times.
For this purpose we use the results of simulations carried out earlier
in \cite{BSD96}. It is to be noted that whereas the simulations
carried out in this paper use a single curve to fit 
the noise power spectrum,
$S_n(f)$, we had used a more accurate representation of $S_n(f)$
in our earlier simulations \cite{BSD96}.
We see that the MLE works moderately well even at low SNRs of
$\rho\approx10$. 
It is to be remarked that the assessment of an estimator depends upon
how we use the estimates to calculate astrophysically interesting
quantities.

\item 
We required about 2
days of compution on a 300 MFlops (peak rating) machine to generate
the results of this paper. 
The use of an integration routine specifically
suited to the integrand, and/or the use of lookup tables for computing
the integrand, would further speed up the computation substantially.
The intrinsic flatness of the manifold turns out to be very convenient
for our purpose. This property holds true for the
first post-Newtonian waveform as well. 
There is one more parameter in the post-Newtonian waveform and
consequently one more integration to perform to get the the marginal
distribution of the parameters.
For higher post-Newtonian corrections and/or for inclusion
of parameters such as spins, it might be computationally expensive to
compute the marginal distributions. However, it is to be noted that
performing Monte Carlo simulations in such cases would also call for a
huge amount of computational effort. 
A further research into the above issues is in progress.
\end{itemize}

\acknowledgements 
R.B. would like to thank  CSIR, India for the senior research
fellowship. S.D. would like to thank Bernard Schutz and Andrjez Krolak
for fruitful discussions.

\begin{figure*}
\caption {Plot of $I(f)$ v/s $f$.}
\label{fig1}
\end{figure*}

\begin{figure*}
\centering
\caption {Variation of root mean square errors in the parameters of the Newtonian
waveform with SNR.}
\label{bsdfig}
\end{figure*}

\begin{figure*}
\centering
\caption {Plot of $\theta_m$ v/s $f$; $\Delta\nu^i\equiv\{1.62, -8.49,
7.3\}$. $\theta_m$ has a value close to $-4\pi$ in the region around 
135Hz, where $I(f)$ attains its maximum.
}
\label{fig2}
\end{figure*}

\begin{figure*}
\centering
\caption {Scatter plot of $\theta_m$; $12000$ realizations of noise have
been considered.}
\label{fig3}
\end{figure*}

\begin{figure*}
\centering
\caption {Scatter plot on the $\nu^1-\nu^2$ plane}
\label{fig4}
\end{figure*}

\begin{figure*}
\centering
\caption {Scatter plot on the $\nu^2-\nu^3$plane }
\label{fig5}
\end{figure*}

\begin{figure*}
\centering
\caption {Schematic illustration of the geometric picture discussed
in the text. 
}
\label{fig6}
\end{figure*}

\begin{figure*}
\centering
\caption {Marginal Porbability distributions of the variables $r^i$.  
}
\label{fig7}
\end{figure*}

\begin{figure*}
\centering
\caption {Plot of the variables $r^i$ v/s $\theta_m$.  
}
\label{fig8}
\end{figure*}

\begin{figure*}
\centering
\caption {Histogram of the variable $\theta_m$.  
}
\label{fig9}
\end{figure*}

\begin{figure*}
\centering
\caption {Comparision between the statistical model and the Monte Carlo
methods to arrive at the marginal distribution of the parameter
$\nu^1$. Part (c) and (d) zoom in to the first and second maxima to
the right of the central maximum. 
}
\label{fig10}
\end{figure*}

\begin{figure*}
\centering
\caption {Histogram of $\theta_m$ for the post-Newtonian case.}
\label{fig13}
\end{figure*}

\begin{figure*}
\centering
\caption {Scatter Plot on the $\nu^1-\nu^2$ plane.}
\label{fig11}
\end{figure*}

\begin{figure*}
\centering
\caption {Scatter Plot on the $\nu^3-\nu^4$ plane.}
\label{fig12}
\end{figure*}

\begin{figure*}
\centering
\caption { Plot of the variables $r^i$ v/s $\theta_m$.}
\label{fig14}
\end{figure*}

\begin{figure*}
\centering
\caption {Distributions of the variables $r^i$ for the post-Newtonian
case are plotted. The histogram is obtained from the Monte-Carlo
simulations whereas the continuuous curve is a unit variance Gaussian curve.}
\label{fig15}
\end{figure*}

\begin{figure*}
\centering
\caption {Plot of $P_{3\sigma}$ for the parameter $\tau_0$
v/s SNR for the Newtonian case.}
\label{fig16}
\end{figure*}
\begin{figure*}
\centering
\caption { Plot of $P_{3\sigma}$ for the parameter $\tau_0$
v/s SNR for the post-Newtonian case.}
\label{fig17}
\end{figure*}
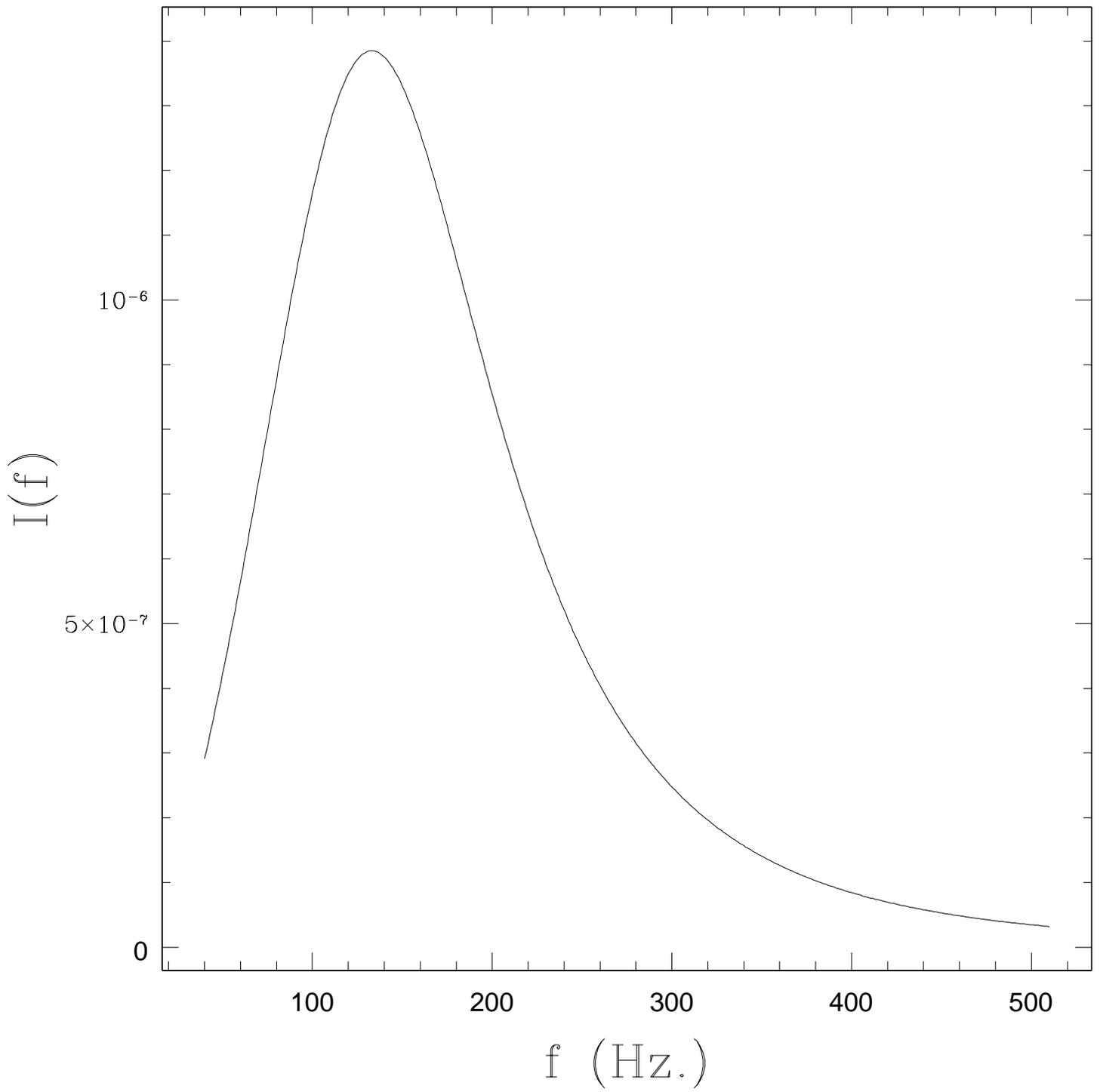
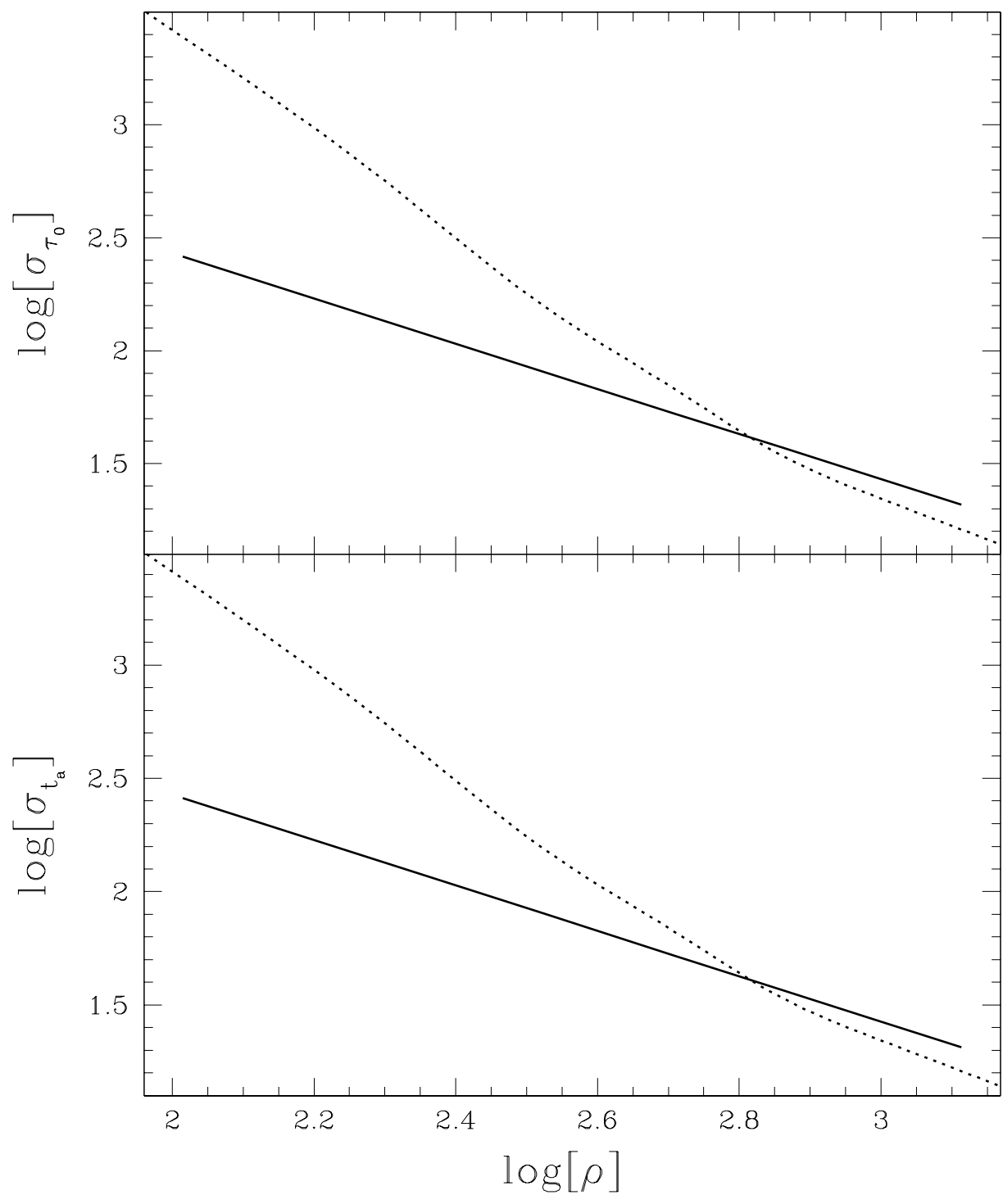
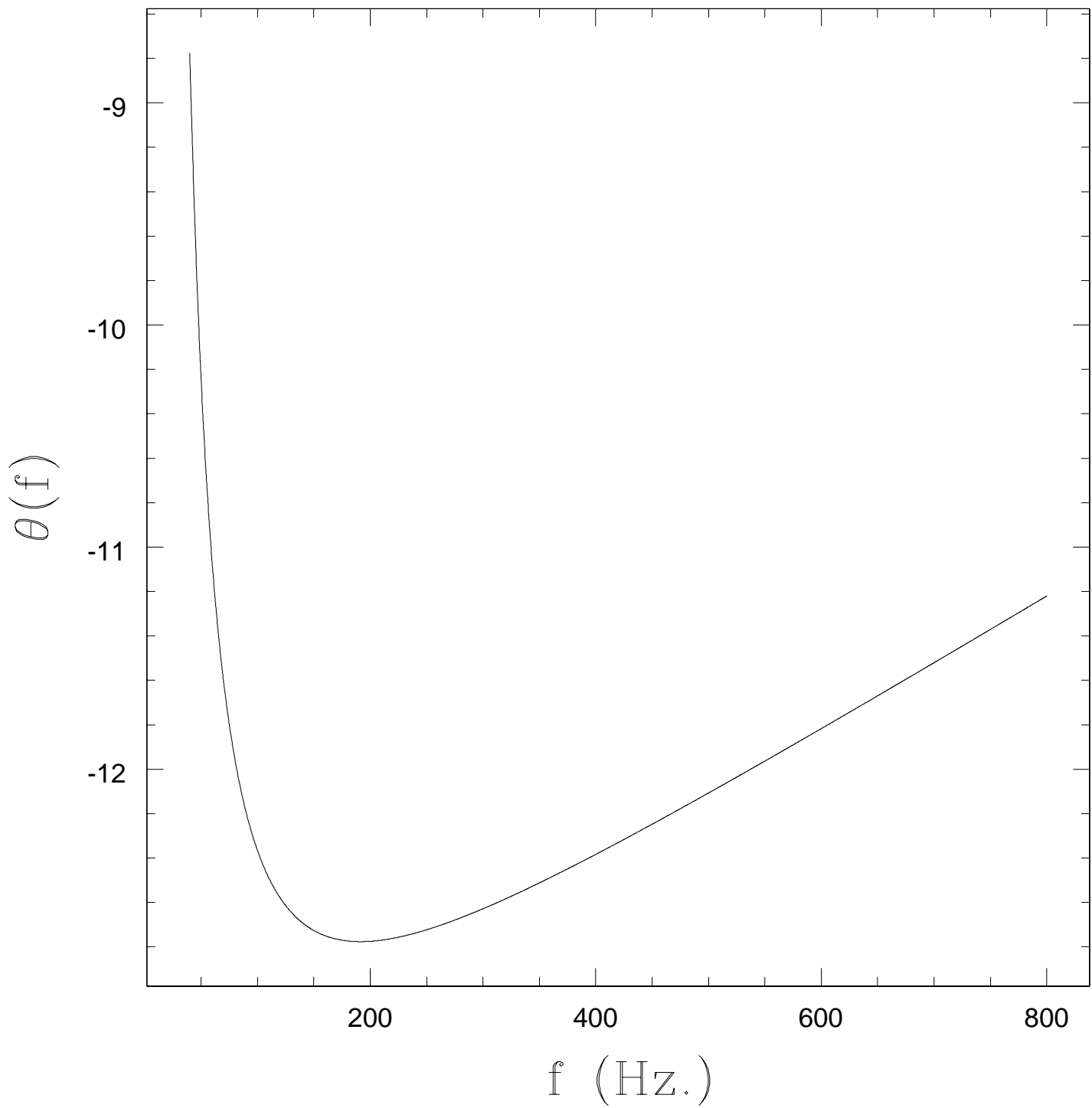
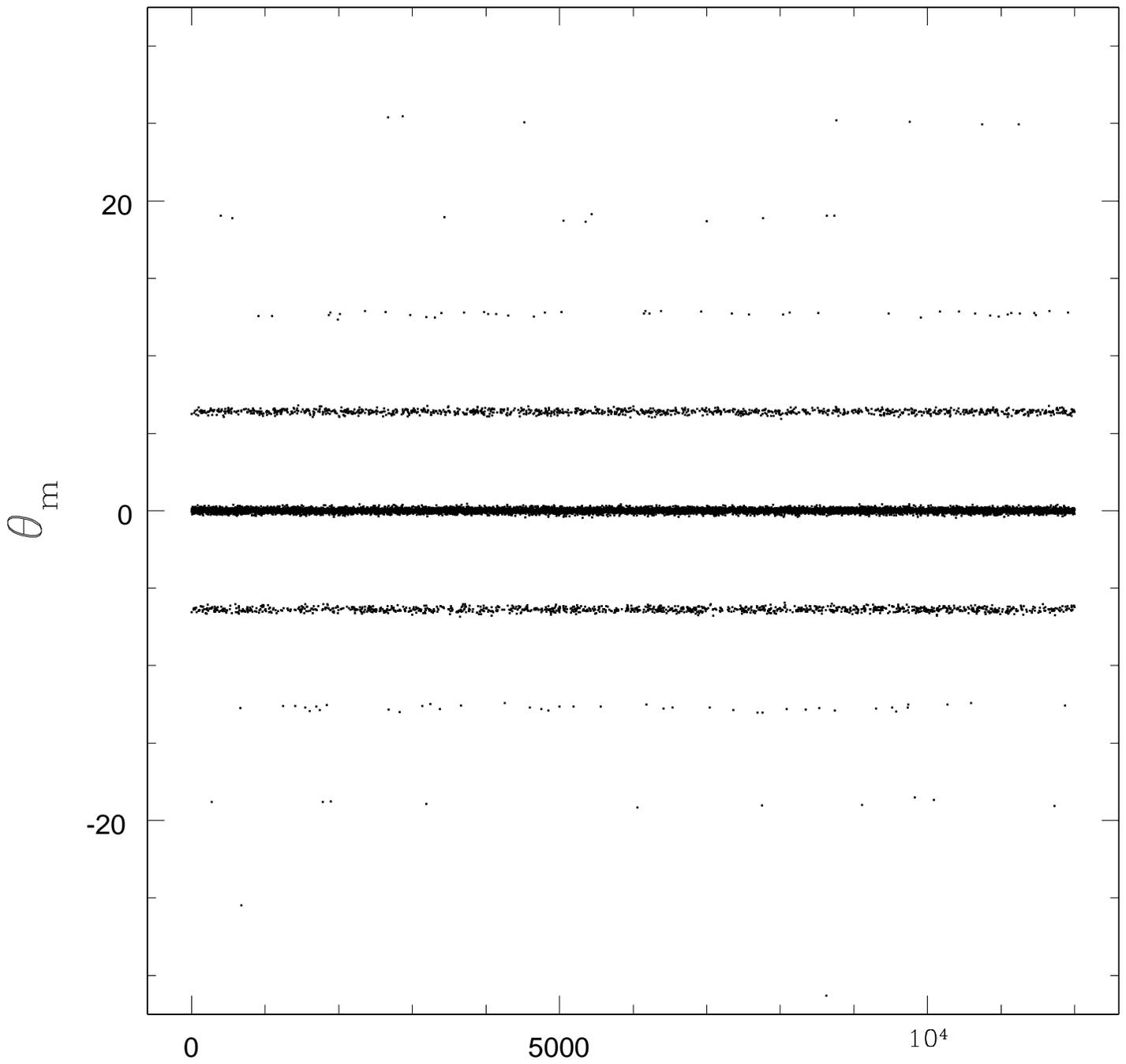
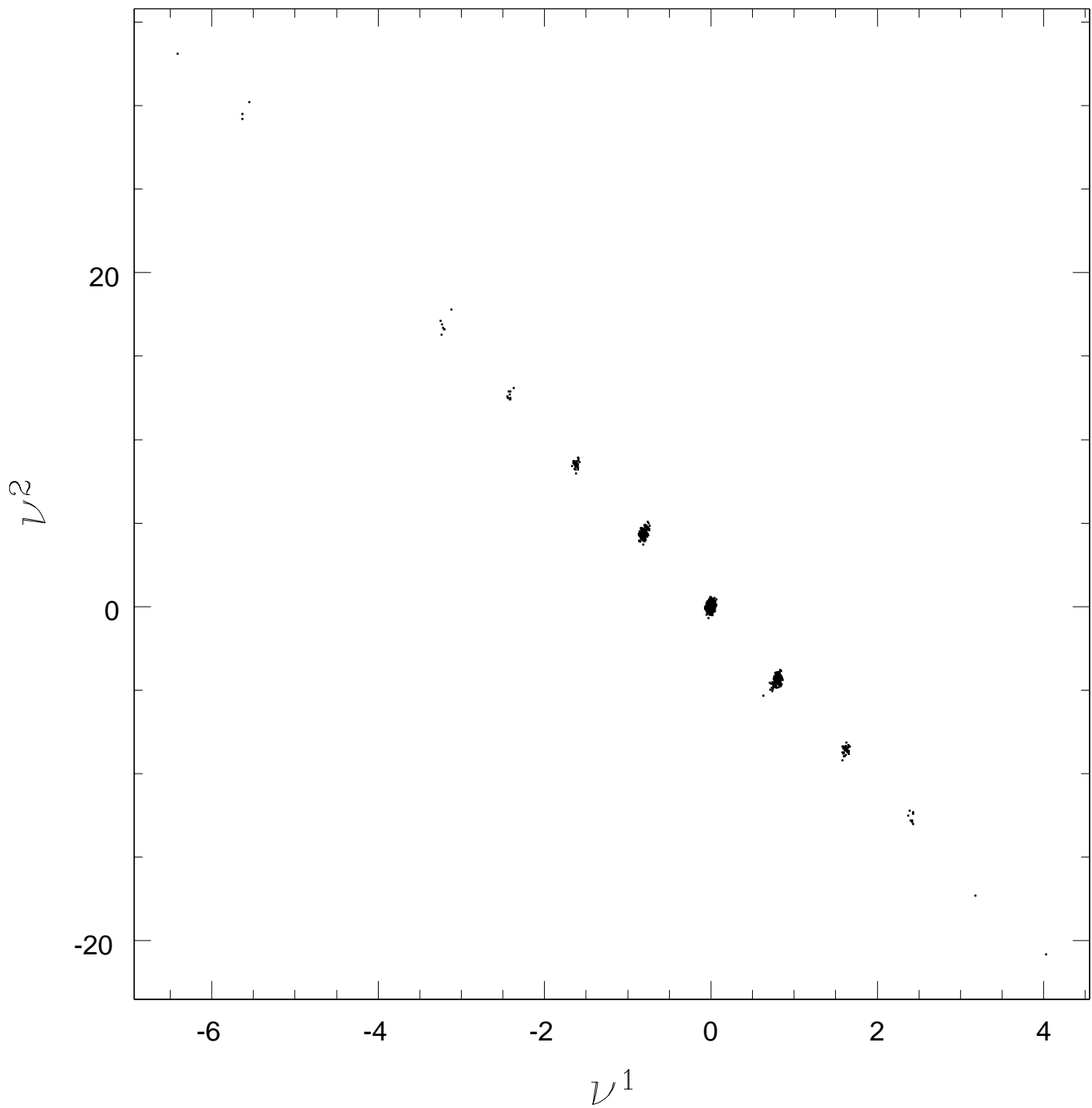
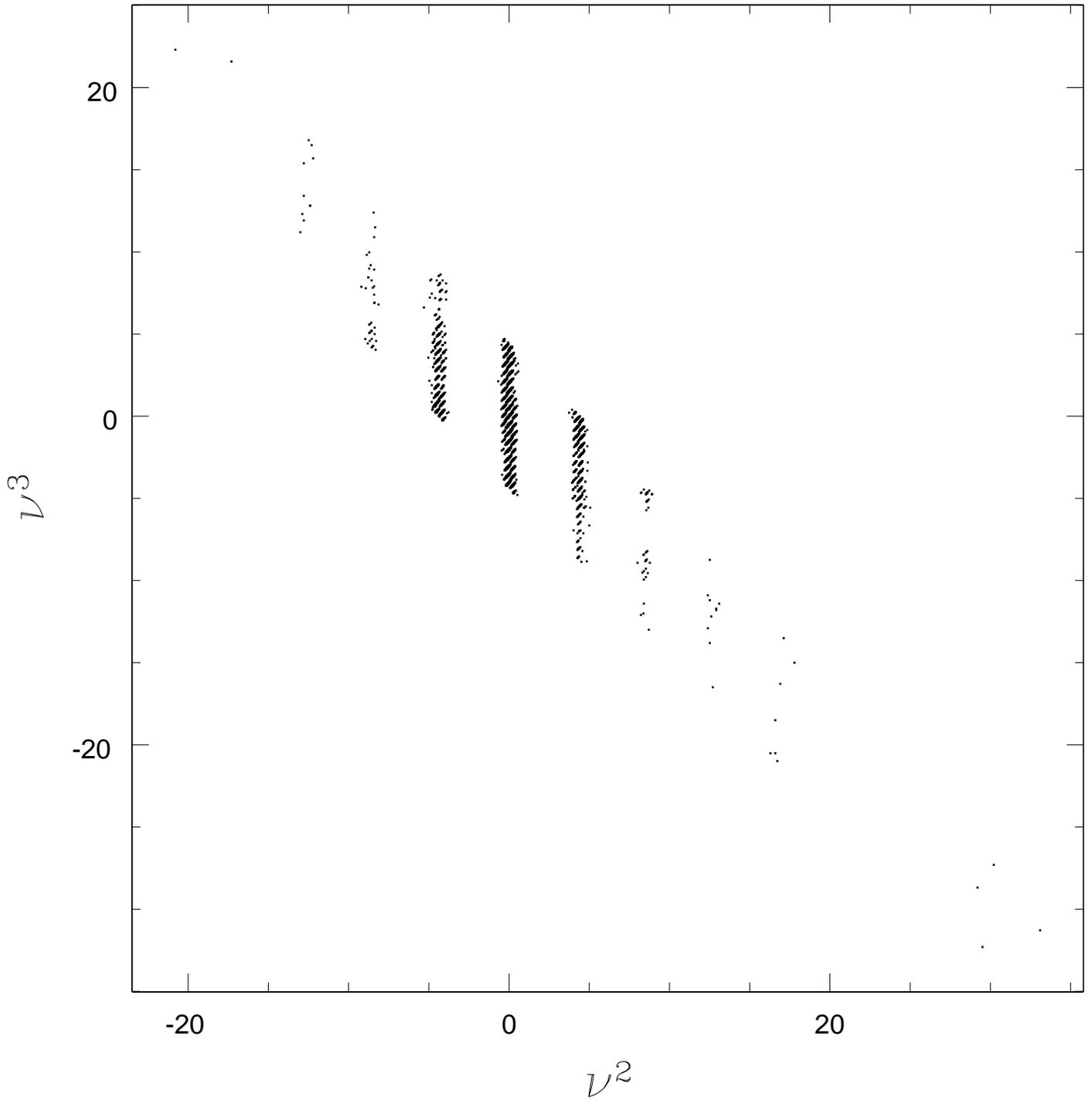
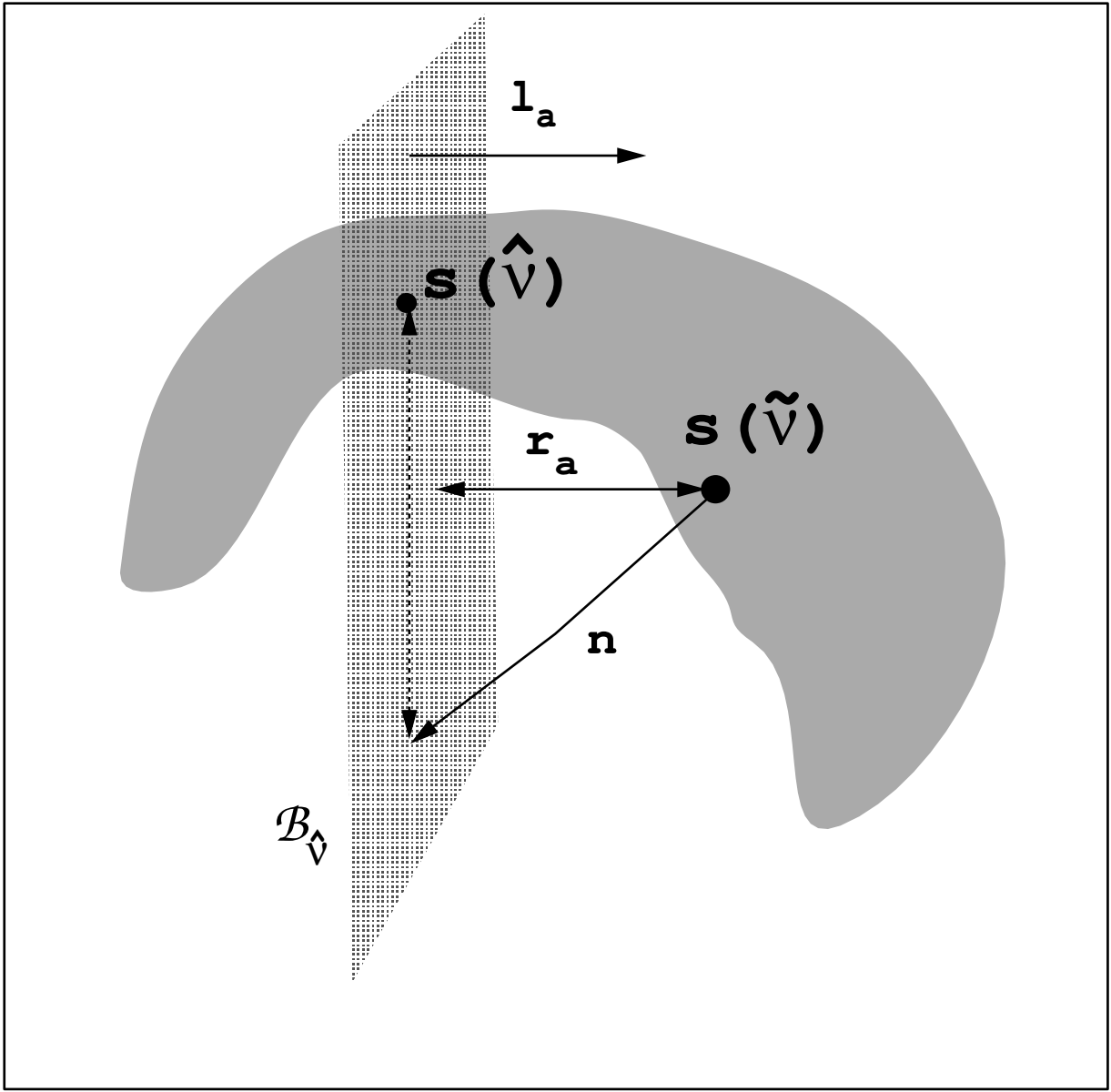
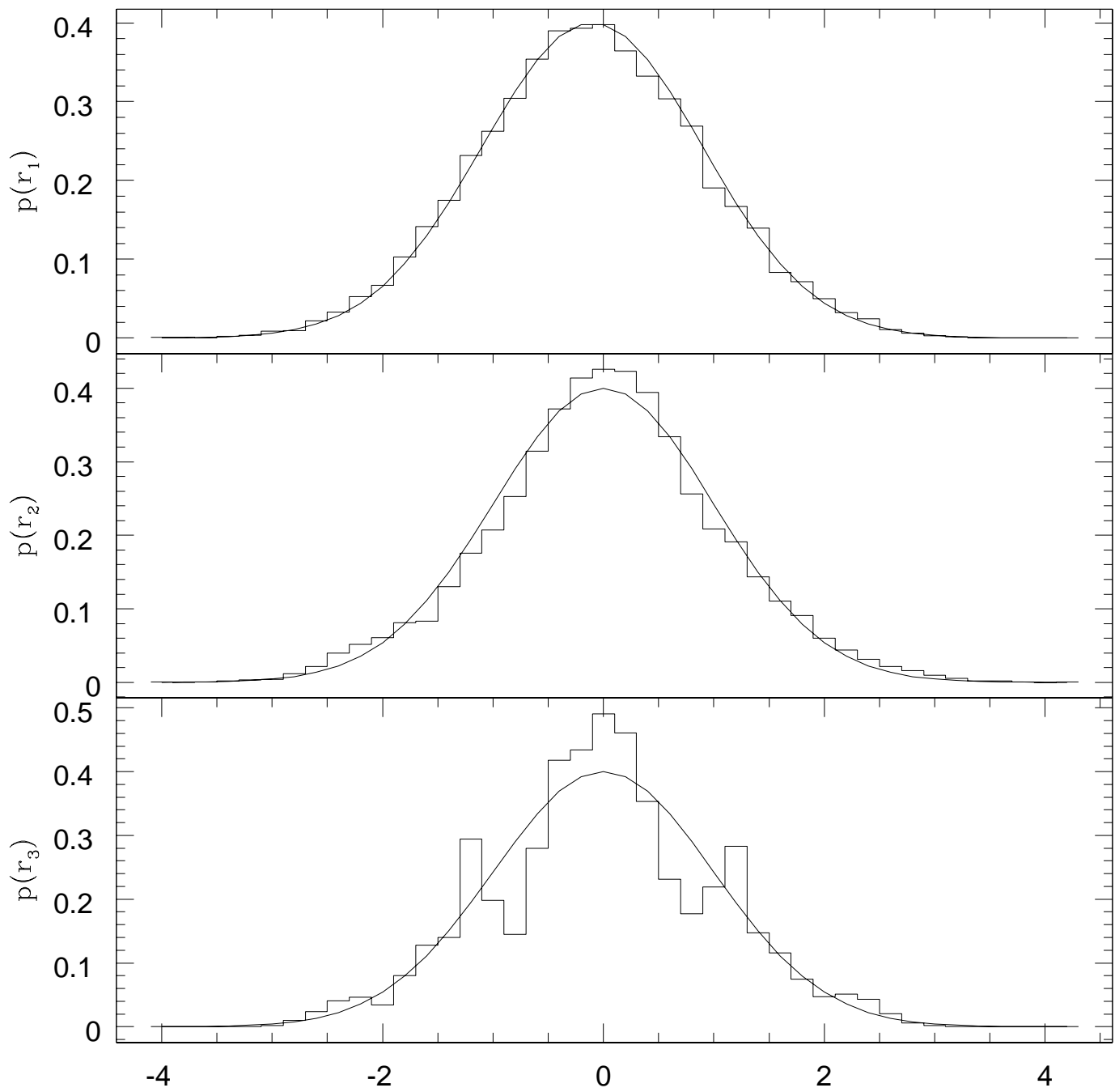
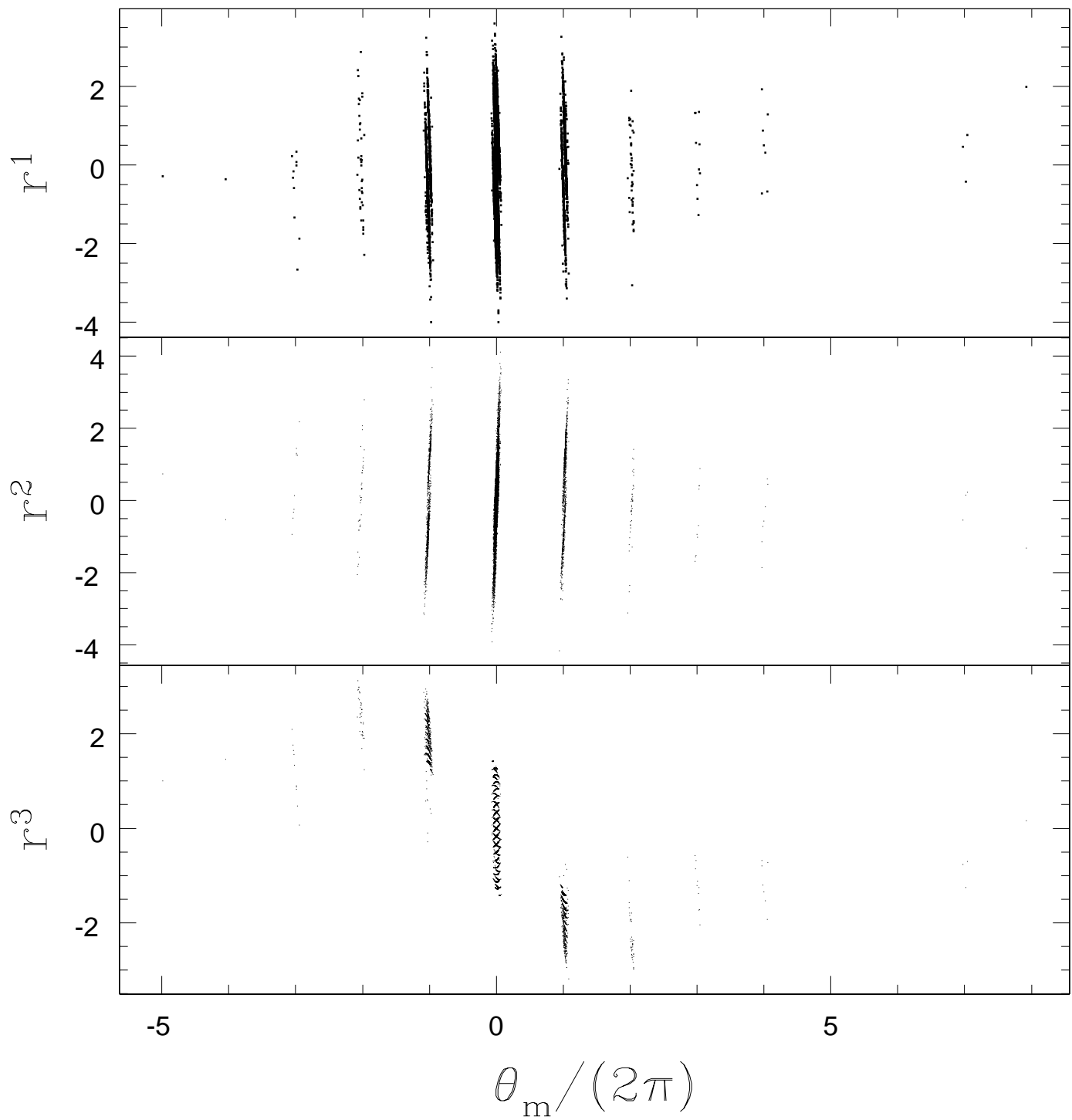
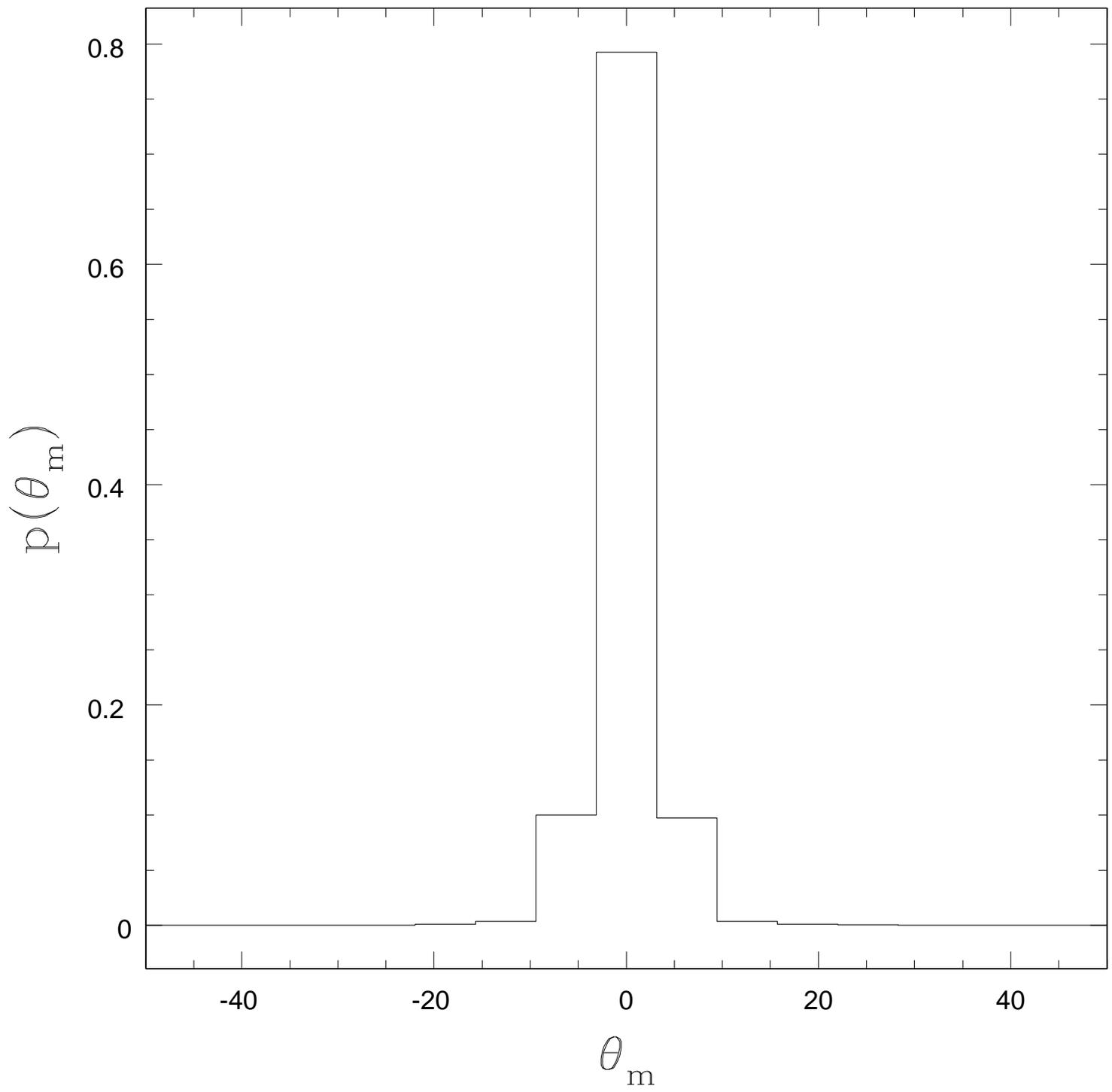
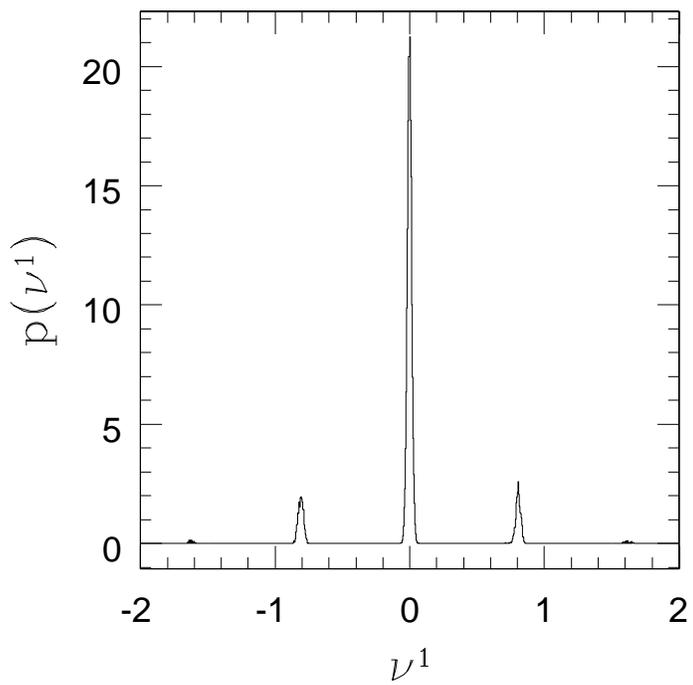
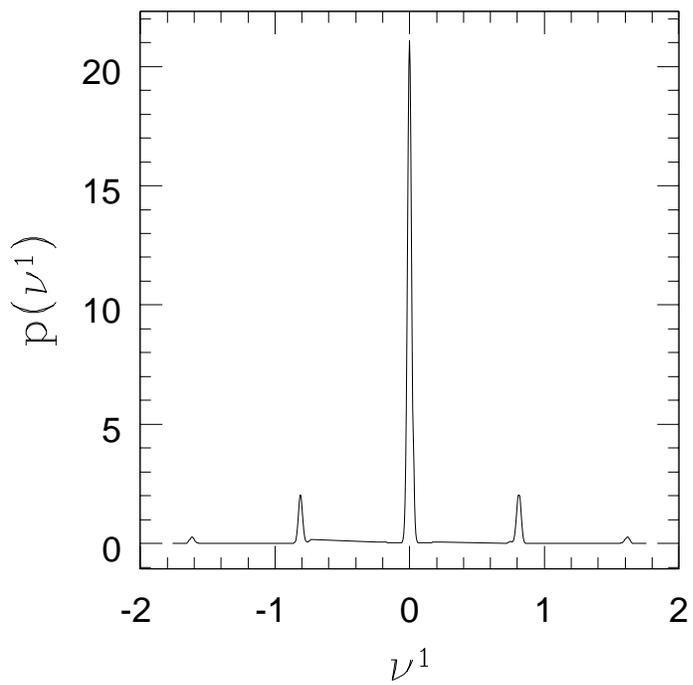
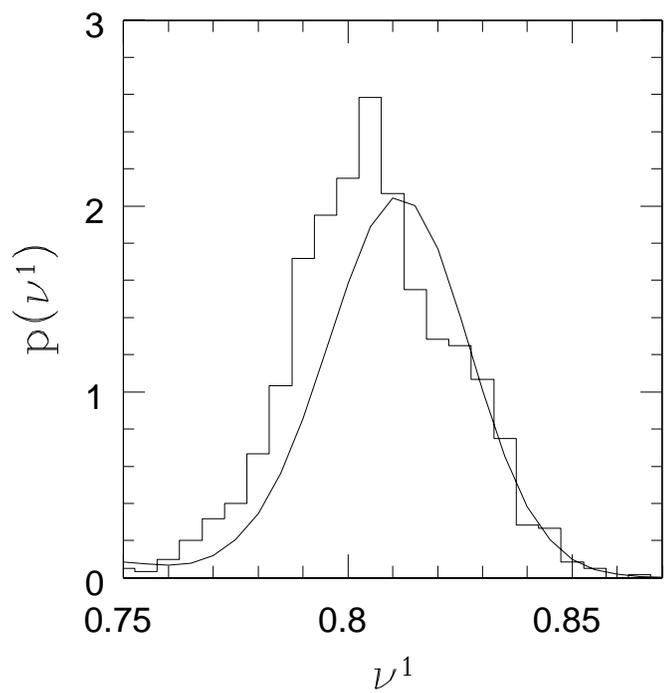
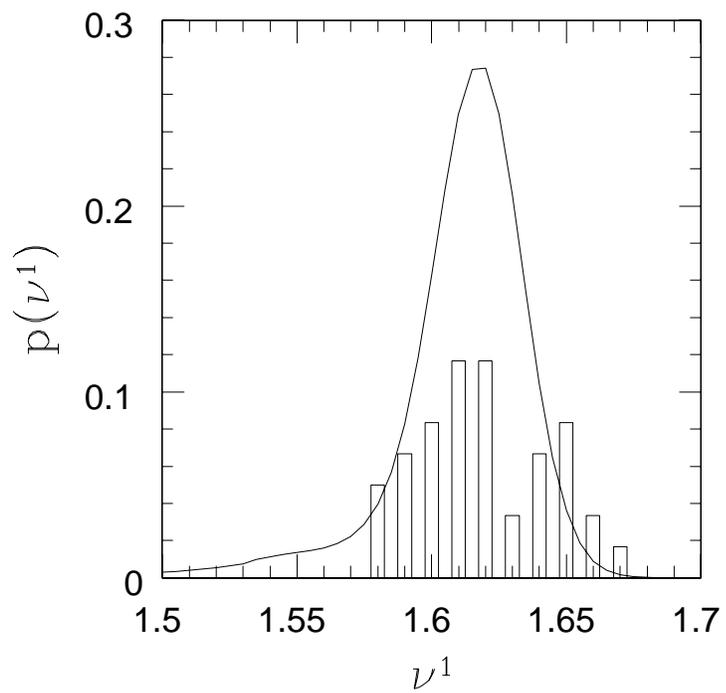
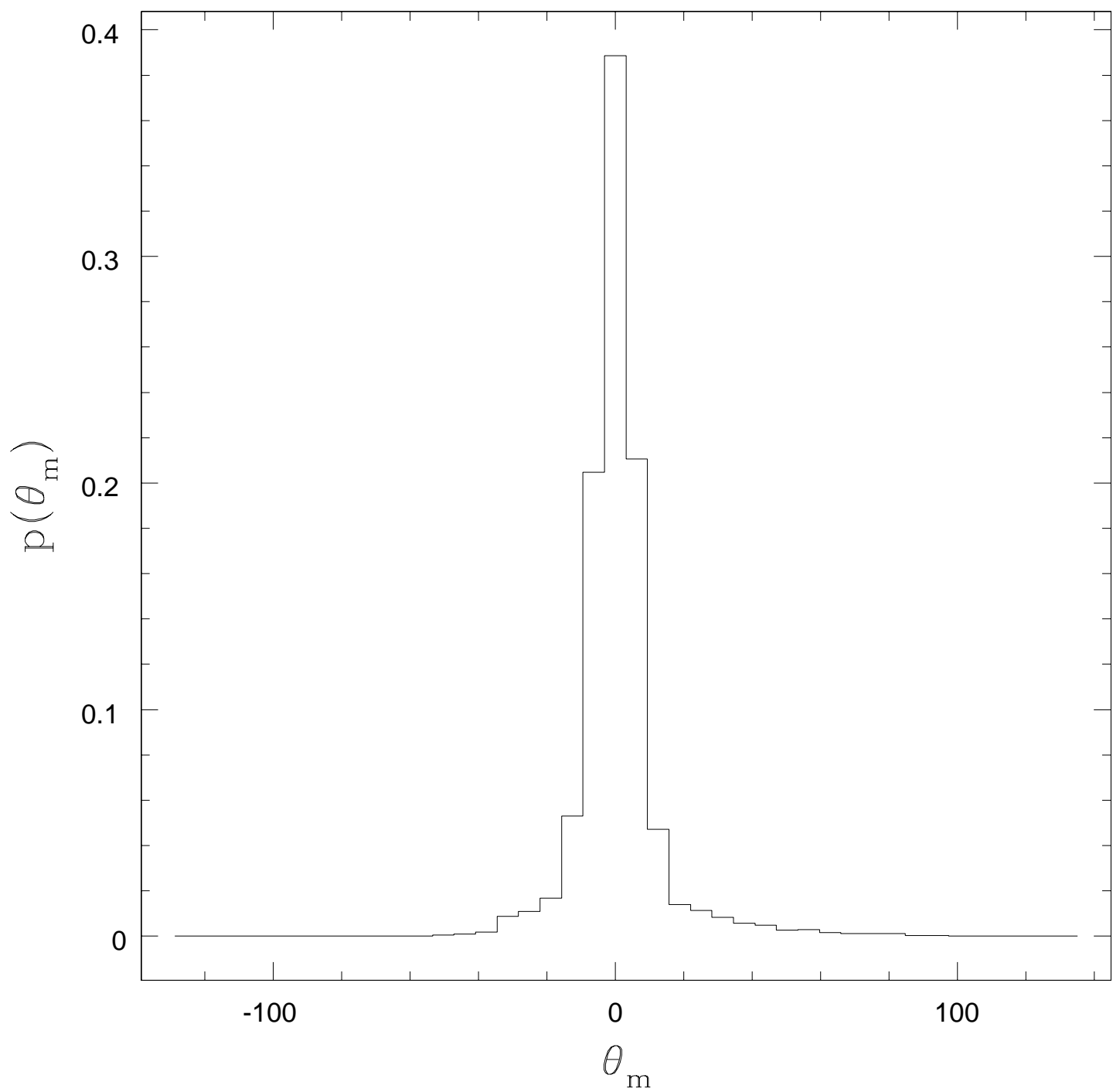
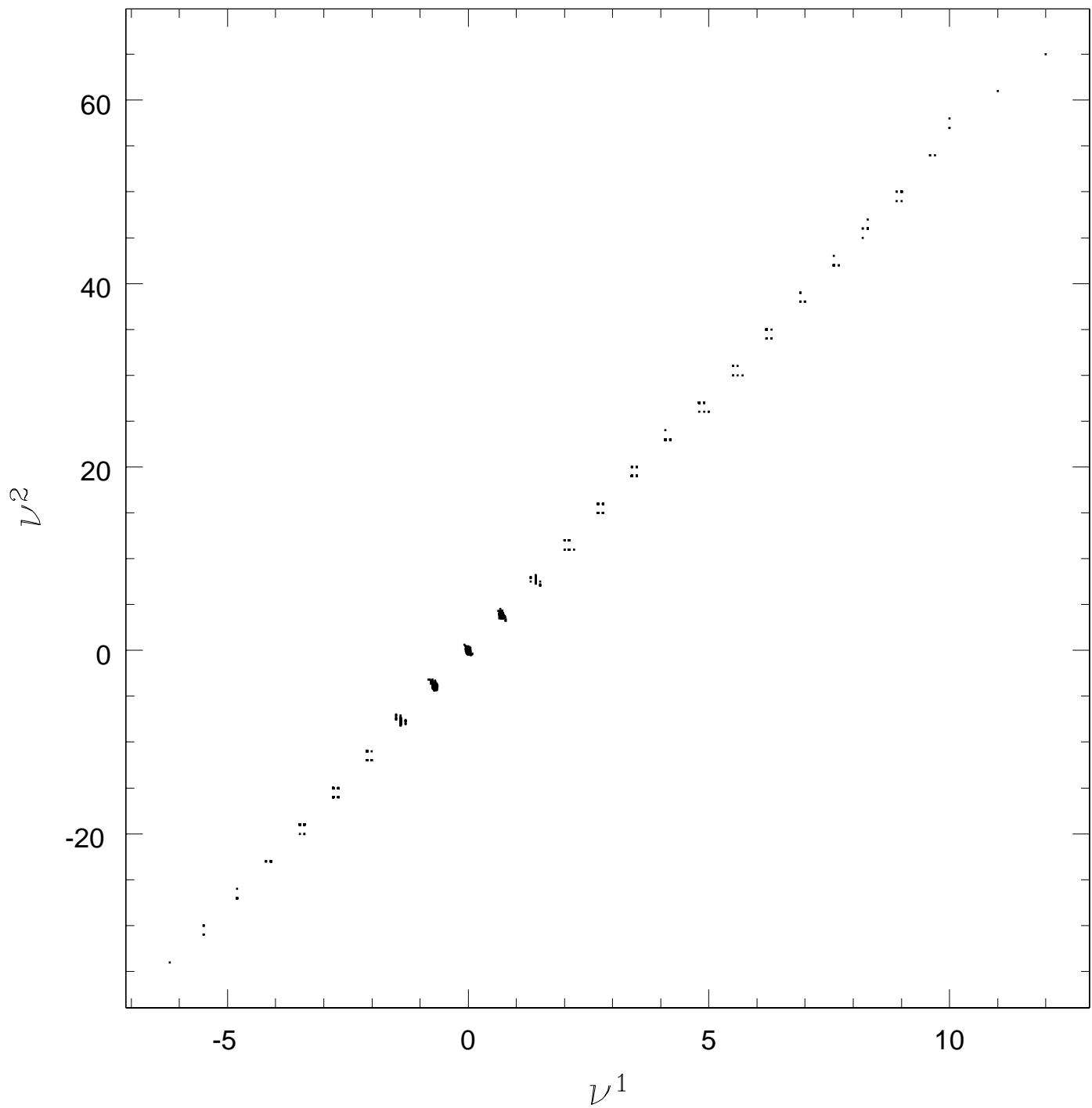
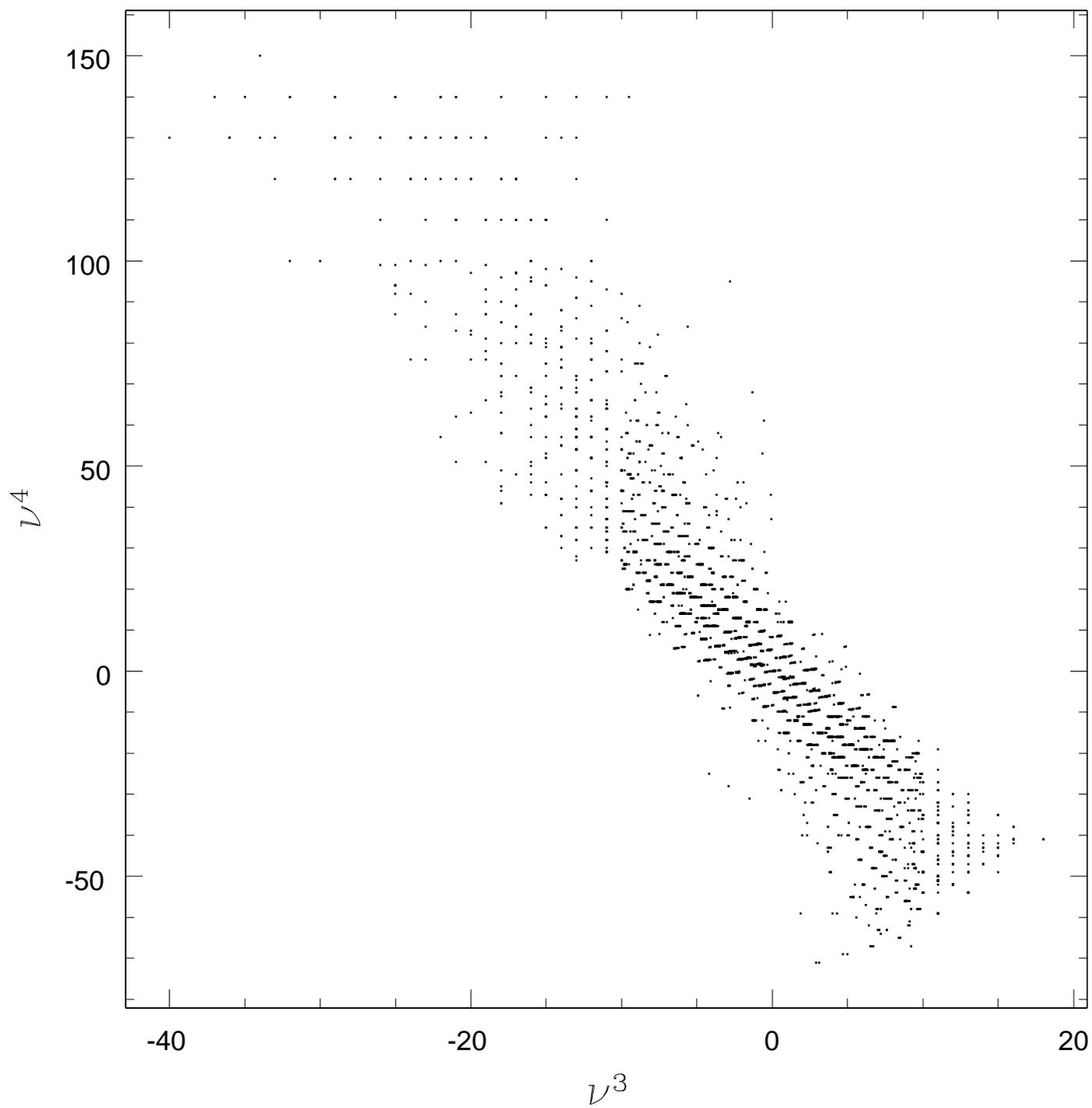
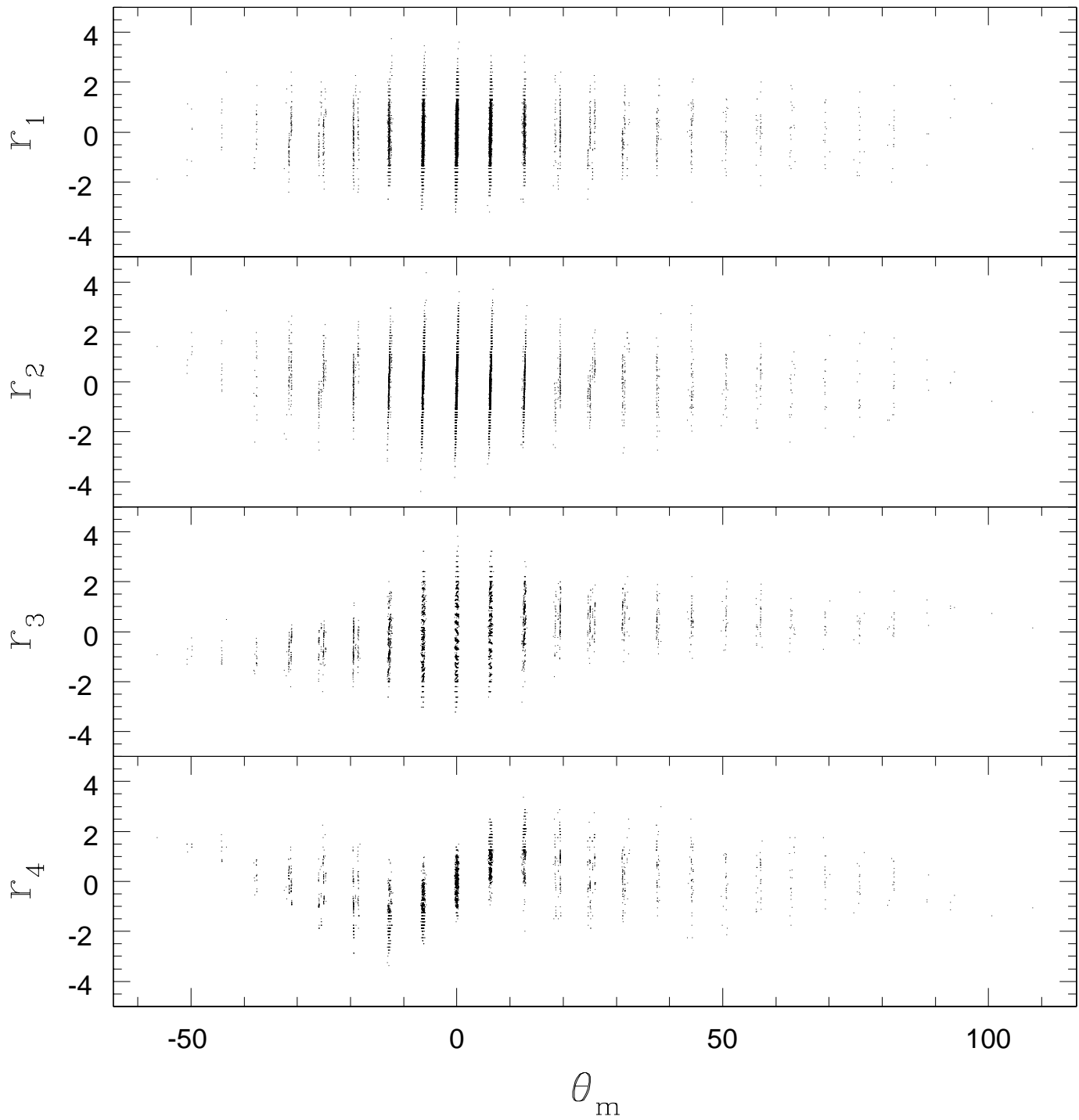
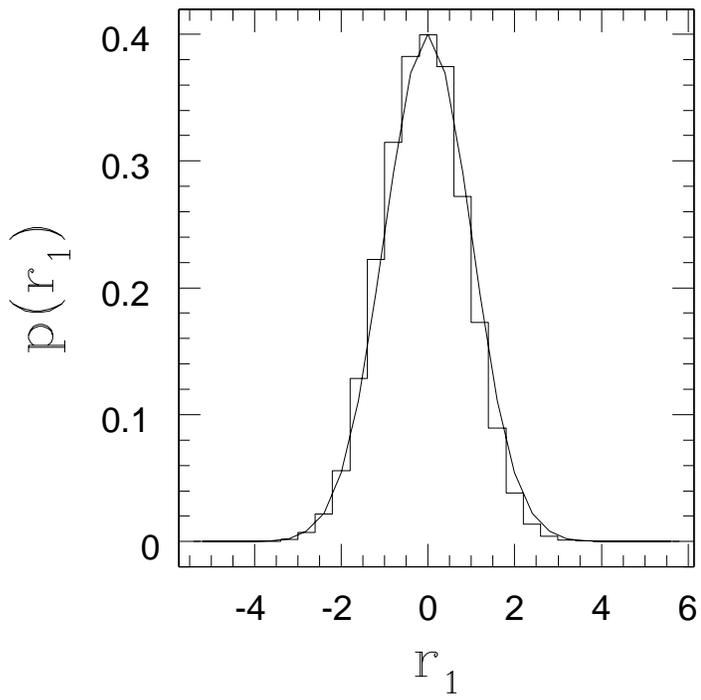
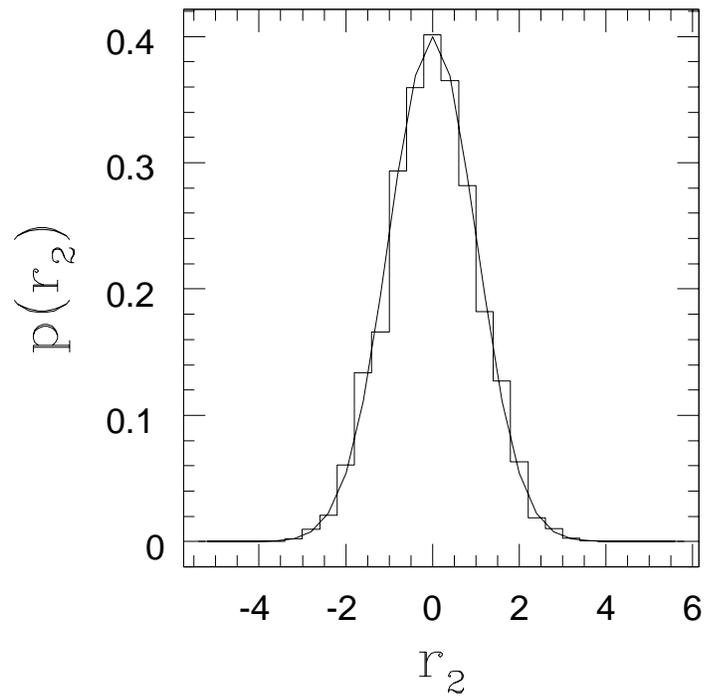
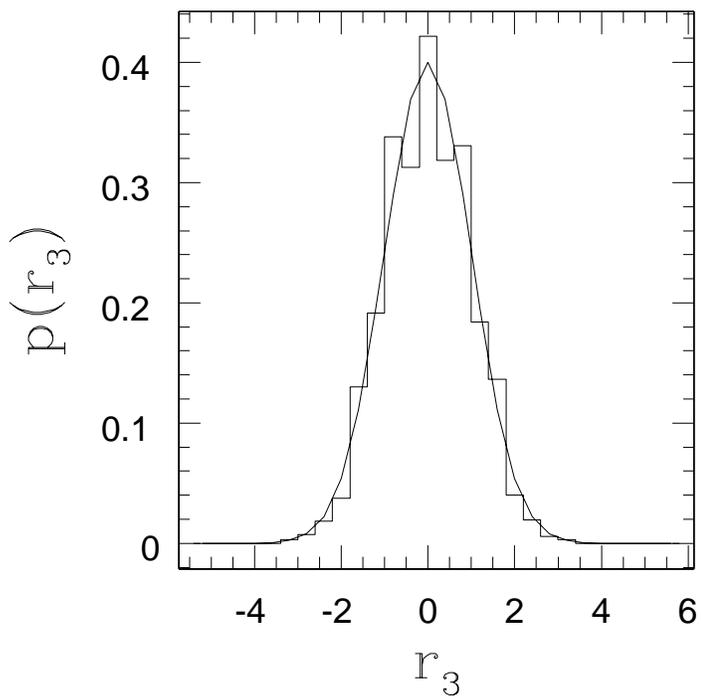
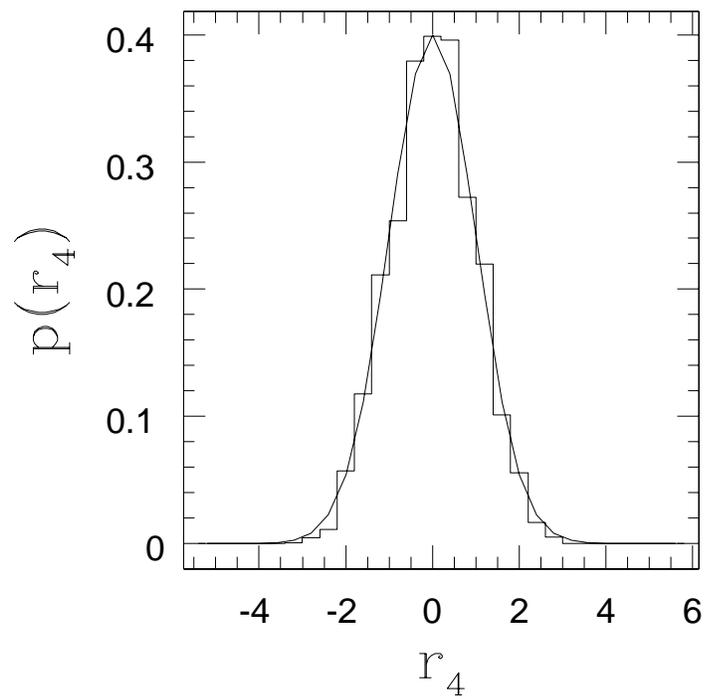
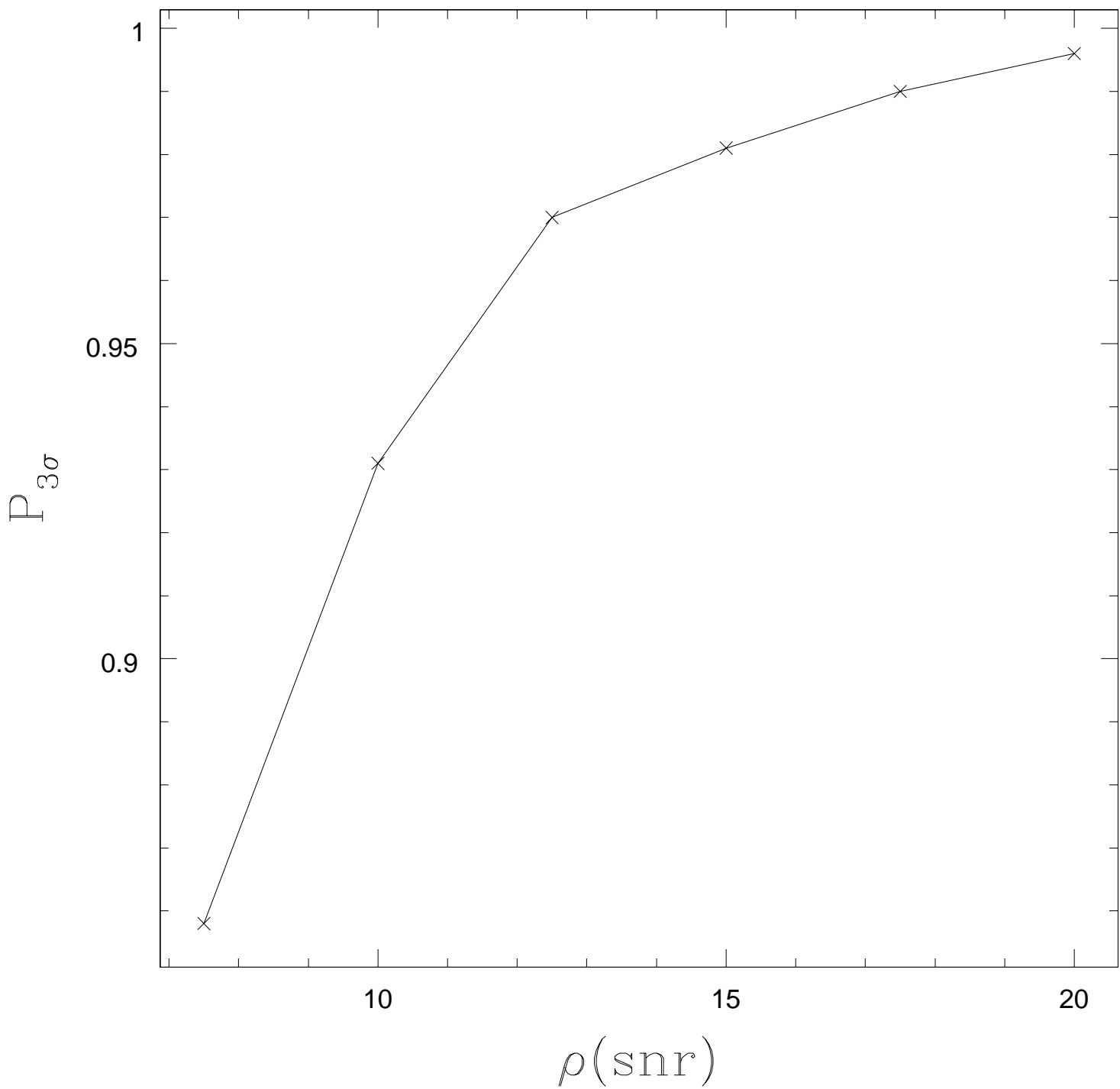
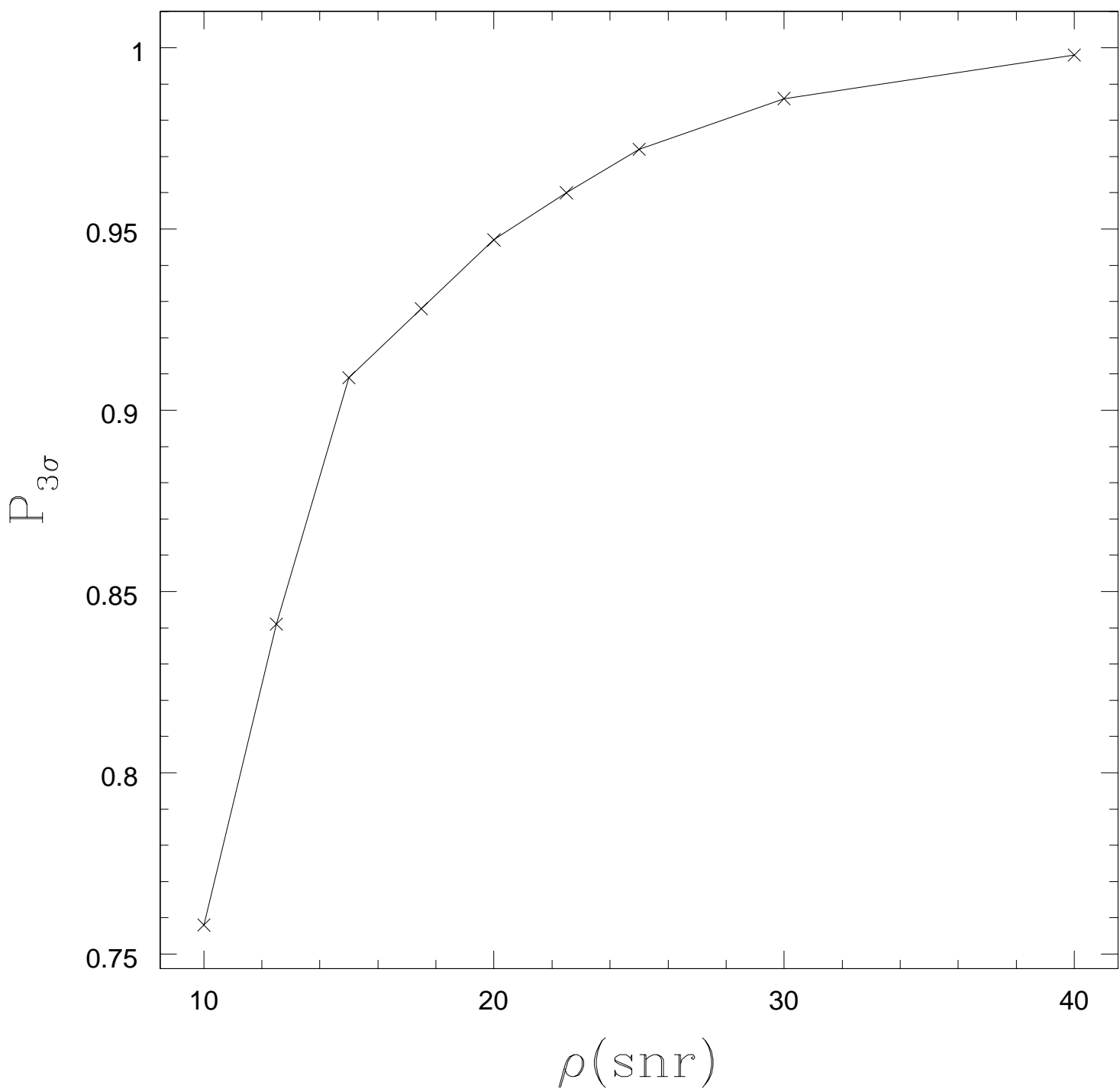
\end{document}